# Black-Box Coupled-Mode Theory: An Ab Initio Framework to Model Electromagnetic Interactions of Open, Lossy, and Dispersive Resonators

Seied Ali Safiabadi Tali[1], Luqi Yuan[2, *], and Wei Zhou[1, *]

[1]Department of Electrical and Computer Engineering, Virginia Tech, Blacksburg, Virginia 24061, United States

[2]State Key Laboratory of Advanced Optical Communication Systems and Networks, School of Physics and Astronomy, Shanghai Jiao Tong University, Shanghai 200240, China

*Corresponding authors: wzh@vt.edu, yuanluqi@sjtu.edu.cn

## Abstract

Temporal coupled-mode theory (TCMT) provides a simple yet powerful platform to model and analyze electromagnetic resonator systems. Nevertheless, restrictive assumptions and lack of rigorous connection to Maxwell's equations limit the TCMT formulation's generality and robustness. Herein, we present the *Black-box Coupled-Mode Theory* (*BBCMT*), a general ab initio CMT framework developed from Maxwell's equations and quasinormal mode (QNM) theory to model the electromagnetic interactions of dispersive, lossy, and open resonators. BBCMT development is enabled by employing Poynting's and conjugated reciprocity theorems to rigorously normalize QNMs and calculate their expansion coefficients. Our novel QNM analysis approach allows the definition of a resonator Hamiltonian matrix to characterize the modes' electromagnetic interactions, energy storage, and absorption. Uniquely, the BBCMT framework can capture a resonator's non-resonant scattering and absorption by treating the resonator scattered fields as a perturbation to those of a background structure, which can be flexibly chosen considering the desired accuracy-simplicity trade-off. BBCMT modeling complexity can be further adjusted by treating user-defined subcomponents of a resonator system as black-boxes described only by their input-output transfer characteristics. Beyond the BBCMT formulation, we present two lemmas to reverse-engineer the modeling parameters from calculated or measured far-field spectra. Moreover, we introduce the signal flow graphs from control theory to illustrate, interpret, and solve the BBCMT equations. To evince BBCMT's validity and generality, we compare the BBCMT and TCMT predictions for two plasmonic nanoresonator systems to finite-difference time-domain simulations and find BBCMT results to be a much better match. BBCMT reduces to TCMT or cavity perturbation theory under certain constraints and approximations.

## 1. Introduction

Modeling and analyzing the scattering and absorption responses of electromagnetic resonator systems is crucial for designing and developing novel devices in antennas, photonics, plasmonics, and metamaterials/metasurfaces [1-7]. Numerical methods, such as finite difference time domain (FDTD), method of moments (MoM), and finite element method (FEM), can precisely solve Maxwell's equations to simulate the electromagnetic responses of resonator systems [8]. Despite providing the far-field and near-field properties of the system, numerical simulations cannot directly interpret the microscopic coupling dynamics between subsystems and interactions of individual modes, constraining their capability for uncovering the underlying design principles. In this regard, complementary analytical methods, such as quasinormal mode (QNM) theory [9-21] and temporal coupled-mode theory (TCMT) [21-29], are required to model and analyze electromagnetic responses of resonator systems in terms of their resonant modes

(generally non-orthogonal). In QNM theory, scattered or total fields created by the action of an electromagnetic source on a resonator are expressed as a linear expansion of its quasinormal modes, i.e., the eigensolutions of sourceless Maxwell's equations with the resonator boundary conditions [19]. The QNM expansion coefficients can then be found using the non-conjugated reciprocity theorem [11, 19] or by residue-decomposition of the resonator dyadic Green's function [16]. Yet, these QNM analysis approaches require normalization of QNMs, which is quite challenging as resonators' QNMs exponentially diverge outside them [9-20]. Two major strategies have been developed to address the QNM normalization issue: 1) Suppressing the diverging QNM fields using perfectly matched layer (PML) boundary conditions and integrating over the whole computational domain, including the PML domain [11, 14, 19]; 2) Integrating over a purely-real finite-domain, without any suppression of QNM fields, and adding surface integral terms to compensate for the exponentially-increasing volume integrals [9, 12, 13, 16]. Despite generating satisfactory results in several particular cases, existing QNM analysis and normalization techniques define the mode overlap integrals using unconjugated field products and arbitrary integration domains. Furthermore, they rely on field-attenuation inside non-physical PML layers (strategy 1) or surface integrals (strategy 2). The mentioned features hinder the establishment of a closed-form relationship between a resonator's energy-related parameters (e.g., total absorption and stored energy) and its QNMs inner products through the application of Poynting's theorem. Since the existing QNM frameworks result in complicated formulations to handle multiple interacting modes in coupled-resonator systems, most QNM studies only focus on uncoupled resonator systems [9-21, 30].

Compared to QNM theory, TCMT has apparent strengths in its formulation simplicity and intuitive microscopic picture to model and analyze both uncoupled and coupled-resonator systems [21-29]. Nonetheless, TCMT formulation is largely phenomenological without a rigorous link to the first-principle Maxwell's equations [22, 25]. Moreover, TCMT formulation is obtained under several restrictive assumptions not applicable in general situations. For instance: 1) The relationship between the input and output coupling coefficients of a resonator's modes is acquired following the time-reversal symmetry assumption, which is only applicable to lossless resonators [22, 24, 25, 27]; 2) Material dispersion effects are neglected [21-28]; 3) The effects of non-orthogonality between a resonator's modes on its absorption and energy storage are neglected [21-25, 27]; 4) Resonators' non-resonant absorption is neglected [21-28]; 5) Despite mentioning the existence of non-resonant scattering in the models, no general approach is offered to acquire it [21-28]; 6) The coupled-resonator formulations are obtained under the assumption of weak coupling between single-mode, closed, and lossless resonators [22, 29]. Therefore, it is desirable to develop a general CMT framework directly from Maxwell's equations that can overcome the constraints and limitations of the current TCMT formulation.

Based on Maxwell's equations and QNM theory, this paper establishes the *Black-box Coupled-Mode Theory* (*BBCMT*), a general ab initio CMT framework to model the electromagnetic interactions of open, lossy, and dispersive resonators. BBCMT development starts with the expansion of a resonator's internally scattered fields in terms of QNMs. However, unlike the conventional QNM analysis approaches, we novelly utilize the conjugated form of Lorentz reciprocity theorem to calculate the QNM expansion coefficients, which allows the definition of mode overlap integrals by volume integrals of conjugated field products over the non-arbitrary integration domain of resonator interior. The combination of our QNM analysis approach with Poynting's theorem enables rigorous normalization of QNMs to define a Hamiltonian matrix for every electromagnetic resonator that, in analogy to quantum mechanics [31], governs the modes electromagnetic interactions, energy storage, and absorption. A key feature in the BBCMT framework is to express a resonator's scattering and absorption as perturbations to a background structure's responses, which can be acquired from analytical, numerical, or experimental data. Background responses can capture the resonator's non-resonant responses, which are especially crucial for modeling the coupled-resonator systems since they modify each resonator's resonance features beyond the effects of mode coupling between the different resonators. The background structure can be freely chosen based on the desired level of accuracy or simplicity, making the BBCMT framework highly flexible. Moreover, the framework can treat user-defined subcomponents of a resonator system as black-boxes for which the input-out transfer characteristics are utilized without involving the internal operation dynamics.

In addition to BBCMT formulation, we introduce the Kappa-estimation and F-estimation lemmas to reverse-engineer the BBCMT modeling parameters from calculated or measured far-field spectra. Furthermore, to better illustrate and interpret the electromagnetic interaction dynamics, we employ the signal flow graphs (SFG) from control engineering, which also turn out to be powerful tools to solve the corresponding BBCMT equations using Mason's theorem [32]. To validate BBCMT's generality, robustness, and flexibility, we present two examples where BBCMT and TCMT predictions are compared to FDTD simulations. The first example uses a metal-insulator-metal (MIM) nanodisk array embedded in a $SiO_2$ slab to study the BBCMT formulation for single resonator systems. The second example utilizes BBCMT formulation for coupled resonators to optimize a nanolaminate plasmonic crystal [33] for the highest absorption performance. In both examples, the results from BBCMT models agree with FDTD simulations much better than the TCMT models. Our analyses show that BBCMT formalism can reduce to TCMT [22-25, 27] and cavity perturbation theory [19, 30, 34] under specific constraints and approximations. Overall, this paper demonstrates that BBCMT can serve as a versatile theoretical framework for modeling, analyzing, and designing electromagnetic resonator systems, especially in areas of plasmonics and metamaterials/metasurfaces.

The paper is organized as follows: In Sec. II, we discuss the BBCMT formulations for a single resonator. In Sec. III, we discuss the loading effects and the BBCMT formulations for coupled-resonator systems. In Sec. IV, we demonstrate BBCMT's validity, generality, and flexibility through a side-by-side comparison of its predictions against TCMT predictions and FDTD simulations. Finally, we conclude our paper in section V.

## 2. BBCMT formulations for a single resonator

This section gives a brief derivation of the general BBCMT formulas for a single resonator and discusses their potential impacts. The detailed derivation process is provided in section 1 of the supplementary information [35]. In subsection 2.1, we first derive the coupled-mode equations for a generic resonator with loss, dispersion, and mode non-orthogonality, and then the equations for the stored energy, scattering, and absorption of the resonator in terms of its mode amplitudes. In subsection 2.2, we discuss the background perturbation for a resonator, which gives a higher level of versatility to the formulations in subsection 2.1. In subsection 2.3, we introduce the concept of ports for a resonator that allows us to turn the equations in subsection 2.2 into the matrix format. In subsection 2.4, we generalize the concept of mode for a resonator by adopting a perturbative approach. In subsection 2.5, we use the formulation obtained in previous subsections to obtain the signal flow graph, characteristic equation, and scattering matrix for a resonator. Finally, in subsection 2.6, we introduce the Kappa-estimation and F-estimation lemmas to reverse-engineer the parameters in a resonator BBCMT model from the scattering and absorption spectra.

### 2.1. Mode amplitudes, energy, scattering, and absorption

We start by considering an arbitrary optical resonator A lying over an arbitrary medium, as shown in Figure 1(a). We use $\varepsilon_{med}(r,\omega)$ and $\varepsilon_A(r,\omega)$ to label the dielectric function when the resonator is absent and the dielectric function when it is present, respectively. $r$ and $\omega$ denote the position and the (angular) frequency, respectively. For simplicity, we assume that both the resonator and medium materials are non-magnetic and isotropic. Suppose that the arbitrary harmonic electromagnetic current sources $J(r,\omega)$ and $M(r,\omega)$ create the incident fields $E^{inc}(r,\omega)$ and $H^{inc}(r,\omega)$ in the absence of resonator A. Moreover, the sources create the total fields $E_A^{tot}(r,\omega)$ and $H_A^{tot}(r,\omega)$ when the resonator is present. We then can define the scattered fields $E_A^{sca}$ and $H_A^{sca}$ as follows [19]:

$$E_A^{sca}(r,\omega) = E_A^{tot}(r,\omega) - E^{inc}(r,\omega), \tag{1a}$$

$$H_A^{sca}(r,\omega) = H_A^{tot}(r,\omega) - H^{inc}(r,\omega). \tag{1b}$$

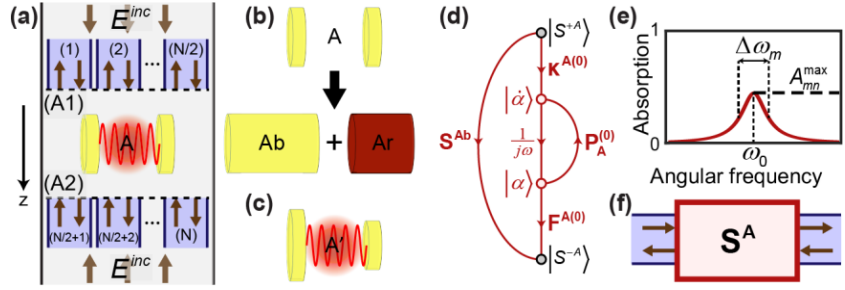

**Figure 1.** BBCMT model for a single resonator. (a) Resonator A is getting excited by the incident field $E^{inc}$. Each waveguide mode in the medium that delivers power to the resonator corresponds to a port. (b) Decomposing a resonator into a background structure (Ab) and a residual structure (Ar). (c) Resonator A' with a geometry close to that of resonator A. (d) The signal flow graph of resonator A showing how its modes are coupled to the input and output power wave vectors. (e, f) Fundamental ideas of Kappa-estimation (e) and F-estimation (f) lemmas to reverse engineer the resonator pole matrix, nonradiative decay rate, and coupling coefficients from the data obtained of numerical simulations or experimental measurements.

The scattered fields represent the resonator response and satisfy [19, 35]:

$$\nabla \times E_A^{sca} = -j\omega\mu_0 H_A^{sca}, \tag{2a}$$

$$\nabla \times H_A^{sca} = j\omega\varepsilon_A E_A^{sca} + j\omega\Delta\varepsilon_A E^{inc}, \tag{2b}$$

where $\Delta\varepsilon_A = \varepsilon_A - \varepsilon_{med}$ and $\mu_0$ is the vacuum permeability. Eqs. (2) indicate that $E^{inc}$ is the source of the scattered fields. We further assume that resonator A's QNMs form a basis set that spans the scattered fields inside it [20, 21], i.e., where $\varepsilon_A \neq \varepsilon_{med}$, and then can expand the scattered fields as:

$$\begin{pmatrix} E_A^{sca}(r,\omega) \\ H_A^{sca}(r,\omega) \end{pmatrix}\bigg|_{r \in \{\varepsilon_A \neq \varepsilon_{med}\}} = \sum_m \tilde{\alpha}_m(\omega) \begin{pmatrix} \tilde{E}_{Am}(r) \\ \tilde{H}_{Am}(r) \end{pmatrix}, \tag{3}$$

where the unitless term $\tilde{\alpha}_m(\omega)$ is the amplitude for the $m^{th}$ resonator mode with the normalized electric and magnetic field patterns of $\tilde{E}_{Am}(r)$ and $\tilde{H}_{Am}(r)$, respectively. We denote the complex resonance frequency of the mode [19, 35] by $\tilde{\omega}_{Am} = \Omega_{Am} + j\Gamma_{Am}$, where $\Omega_{Am}$ and $\Gamma_{Am}$ determine the mode oscillation frequency and (total) decay rate, respectively. By definition, the field patterns and the resonance frequency of the $m^{th}$ mode satisfy the following set of eigenequations:

$$\nabla \times \tilde{E}_{Am}(r) = -j\tilde{\omega}_{Am}\mu_0 \tilde{H}_{Am}(r), \tag{4a}$$

$$\nabla \times \tilde{H}_{Am}(r) = j\tilde{\omega}_{Am}\varepsilon_A(r,\tilde{\omega}_m)\tilde{E}_{Am}(r). \tag{4b}$$

We choose the normalization condition for the $m^{th}$ mode as follows:

$$\tilde{W}_{Am} = \frac{1}{4} \iiint_{\varepsilon_A \neq \varepsilon_{med}} \left[ \text{Re}\left(\frac{\partial(\omega\varepsilon_A)}{\partial\omega}\right)_{\omega=\Omega_{Am}} |\tilde{E}_{Am}|^2 + \mu_0 |\tilde{H}_{Am}|^2 \right] d^3r = 1. \tag{5}$$

The integration in Eq. (5) can be done over one unit cell if the resonator has a periodic structure, e.g., for a nanohole array [36]. $|\tilde{\alpha}_m|^2 \tilde{W}_{Am}$ can be interpreted as the time-averaged electromagnetic energy stored in the $m^{th}$ mode [37]. By applying the Poynting's theorem to Eqs. (4) [19, 35], we can decompose the (total) decay rate of the $m^{th}$ mode as $\Gamma_{Am} = \Gamma_{Am}^r + \Gamma_{Am}^{nr}$, where $\Gamma_{Am}^r = 1/\tau_{Am}^r$ and $\Gamma_{Am}^{nr} = 1/\tau_{Am}^{nr}$ are the radiative and nonradiative decay rates, respectively. $\tau_{Am}^r$ and $\tau_{Am}^{nr}$ are the corresponding radiative and nonradiative lifetimes.

We now derive the steady-state coupled-mode (CM) equation for resonator A. By using the general form of the conjugated Lorentz reciprocity theorem [11] together with Eqs. (2), (3) and (4), we obtain [35]:

$$\mathbf{H}_A(\omega) \begin{bmatrix} j(\omega-\tilde{\omega}_1)\tilde{\alpha}_1 \\ j(\omega-\tilde{\omega}_2)\tilde{\alpha}_2 \\ \vdots \end{bmatrix} = -\frac{j\omega}{4} \iiint_{\varepsilon_A \neq \varepsilon_{med}} \Delta\varepsilon_A \cdot E^{inc} \cdot \begin{bmatrix} \tilde{E}_{A1}^* \\ \tilde{E}_{A2}^* \\ \vdots \end{bmatrix} d^3r, \tag{6}$$

where the $\mathbf{M} \times \mathbf{M}$ matrix $\mathbf{H}_A$ is the *Hamiltonian matrix* with $\mathbf{M}$ being the number of modes in resonator A. The elements of $\mathbf{H}_A$ are defined as follows:

$$h_{A,mn} = \frac{1}{4}(\omega-\tilde{\omega}_{Am})^{-1} \iiint_{\varepsilon_A \neq \varepsilon_{med}} \left\{ [\omega\varepsilon_A(\omega) - \tilde{\omega}_{Am}\varepsilon_A(\tilde{\omega}_{Am})](\tilde{E}_{Am}.\tilde{E}_{An}^*) + [\omega\mu_0 - \tilde{\omega}_{Am}\mu_0](\tilde{H}_{Am}.\tilde{H}_{An}^*) \right\} d^3r \approx$$

$$\frac{1}{4} \iiint_{\varepsilon_A \neq \varepsilon_{med}} \left\{ \frac{\partial[\omega\varepsilon_A(\omega)]}{\partial\omega} \bigg|_{\omega=\tilde{\omega}_{Am}} (\tilde{E}_{Am}.\tilde{E}_{An}^*) + \mu_0(\tilde{H}_{Am}.\tilde{H}_{An}^*) \right\} d^3r \approx$$

$$\frac{1}{4} \iiint_{\varepsilon_A \neq \varepsilon_{med}} \left\{ \text{Re}\left[\frac{\partial[\omega\varepsilon_A(\omega)]}{\partial\omega}\right]_{\omega=\Omega_{Am}} (\tilde{E}_{Am}.\tilde{E}_{An}^*) + \mu_0(\tilde{H}_{Am}.\tilde{H}_{An}^*) \right\} d^3r \tag{7}$$

$$-\frac{j}{4} \iiint_{\varepsilon_A \neq \varepsilon_{med}} \text{Im}\left[\frac{\partial[\omega\varepsilon_A(\omega)]}{\partial\omega}\right]_{\omega=\Omega_{Am}} (\tilde{E}_{Am}.\tilde{E}_{An}^*) d^3r.$$

Eq. (7) indicates that the Hamiltonian matrix is generally non-Hermitian and its elements have the unit of energy. When the resonator materials are low-dispersion, i.e., $|\partial(\omega\varepsilon_A)/(\varepsilon_A \partial\omega)| \approx 1$, the total energy stored in the resonator modes would be:

$$W_A = \frac{1}{4} \iiint_{\varepsilon_A \neq \varepsilon_{med}} \left[ \text{Re}\left(\frac{\partial(\omega\varepsilon)}{\partial\omega}\right) \left|\sum_m \tilde{\alpha}_m \tilde{E}_{Am}\right|^2 + \mu_0 \left|\sum_m \tilde{\alpha}_m \tilde{H}_{Am}\right|^2 \right] d^3r = \text{Re}\left[\langle\tilde{\alpha}|\mathbf{H_A}|\tilde{\alpha}\rangle\right], \qquad (8)$$

where $|\alpha\rangle$ is the vector consisting of the amplitude terms $\alpha_m$ for all the modes. In the special case that the resonator modes are orthogonal, i.e., $h_{mn} \approx 0$ for $m \neq n$, we can use Eq. (5) to further simplify Eq. (7) as:

$$h_{mn} \approx \left(1 - \frac{j\Gamma_{Am}^{nr}}{\Omega_{Am}}\right)\delta_{mn}, \qquad (9)$$

With Eq. (9), the total energy in Eq. (8) becomes $W_A = \langle\tilde{\alpha}|\tilde{\alpha}\rangle$, which is consistent with the formula of stored energy in TCMT [22, 24, 25].

To calculate the scattered fields of resonator A, we use the Lippmann-Schwinger equation for electrodynamics [20]. By rewriting the right-hand side of Eq. (2b) as $j\omega\varepsilon_{med}E_A^{sca} + j\omega\Delta\varepsilon_A\left(E^{inc} + E_A^{sca}\right)$, and using Eq. (3), we get the following expression for the scattered electric field [35]:

$$E_A^{sca} = E_{Af}^x + \sum_m \tilde{\alpha}_m \tilde{E}_{Am}^{sca}, \qquad (10)$$

where $E_{Af}^x(r,\omega) = \omega^2\mu_0 \iiint_{\varepsilon_A \neq \varepsilon_{med}} G_{med}^E \cdot \Delta\varepsilon_A \cdot E^{inc} d^3r'$ and $\tilde{E}_{Am}^{sca}(r,\omega) = \omega^2\mu_0 \iiint_{\varepsilon_A \neq \varepsilon_{med}} G_{med}^E \cdot \Delta\varepsilon_A \cdot \tilde{E}_{Am} d^3r'$ with $G_{med}^E(r,r',\omega)$ being the electric dyadic Green's function of the medium. $E_{Af}^x$ represents the non-resonantly scattered electric field (not mediated by the modes) which is an essential aspect of the Fano effect [21, 24, 28]. $\tilde{E}_{Am}^{sca}$ denotes the normalized scattered electric field for the $m^{th}$ mode. In general, $\tilde{E}_{Am}^{sca}(r,\omega) \neq \tilde{E}_{Am}(r)$. However, by writing the Lippmann-Schwinger equation based on Eq. (4), we can obtain that $\tilde{E}_{Am}^{sca}(r,\tilde{\omega}_m) = \tilde{E}_{Am}(r)$. Similar expressions can be found for the scattered magnetic field $H_A^{sca}$.

If the resonator materials are low-dispersion and low-loss, i.e., $\tan(\delta_A) = -\text{Im}[\varepsilon_A]/\text{Re}[\varepsilon_A] \ll 1$, we can write the resonator absorption (dissipated power) as [35]:

$$A_A(\omega) \approx A_{Af}^x - 2\omega\text{Im}\left[\langle\tilde{\alpha}|\mathbf{H_A}|\tilde{\alpha}\rangle\right], \qquad (11)$$

with $A_{Af}^x(\omega) = -\frac{\omega}{2} \iiint_{\varepsilon_A \neq \varepsilon_{med}} \text{Im}(\varepsilon_A)|E^{inc}|^2 d^3r$ being the non-resonant absorption. From Eq. (9), we can simplify Eq. (11) to $A_A \approx A_{Af}^x + 2\omega\langle\tilde{\alpha}|\mathbf{\Gamma_A^{nr}\Omega_A^{-1}}|\tilde{\alpha}\rangle$ under the mode orthogonality condition. If the modes are also spectrally narrow (quality factors of 10 or above), Eq. (11) further reduces to $A_A = A_{Af}^x + \langle\tilde{\alpha}|2\mathbf{\Gamma_A^{nr}}|\tilde{\alpha}\rangle$.

## 2.2. Resonator background perturbation

While the non-resonant scattering and absorption terms in Eqs. (10) and (11) can be calculated analytically, they are difficult to be extracted from numerical simulations or experimental measurements. The common method to alleviate this issue is assuming a specific non-resonant spectral background, which typically requires extra assumptions not applicable in the general case [21, 24, 27, 28], such as the assumptions that the resonator dimensions are much smaller than the wavelength [28], the resonator has a mirror symmetry in the z-direction [24], and the materials have a large refractive index [21, 27].

Another issue with Eqs. (10) and (11) is that they involve all the resonator modes. As a resonator usually has an infinite number of modes, it is impossible to account for all of them due to limited computational resources. We are usually only interested in the resonator's scattering and absorption responses in a specific frequency range and from some principal modes active in some principal areas of interest in the resonator (e.g., cavities, protrusions, etc.). Modes that are far from resonance typically have a negligible contribution to the resonator's total scattering and absorption. However, the resonator's non-principal modes (those having small fields in the principal areas) may collectively have a non-negligible contribution if they are close to resonance. For example, in a metal-insulator-metal nanoantenna with a gap as the principal area, we are exclusively interested in the scattering response from the gap modes while the electric dipole mode also gives a considerable scattering response, despite not being active in the gap [38].

However, as we will show in the following, the issues mentioned above can be alleviated using a perturbational analysis. Consider a structure with the dielectric function $\varepsilon_{Ab}(r,\omega)$ such that the region with $\varepsilon_A \neq \varepsilon_{Ab}$ coincides with the principal areas in the resonator A (Figure 1(b)). We call the structure corresponding to $\varepsilon_{Ab}(r,\omega)$ the *background structure* of resonator A. The scattered electric field of resonator A can then be decomposed as:

$$E_A^{sca} = E_{Ab}^{sca} + E_{Ar}^{sca}, \tag{12}$$

where $E_{Ab}^{sca} = \omega^2 \mu_0 \iiint_{\varepsilon_{Ab} \neq \varepsilon_{med}} G_{Ab}^E . \Delta \varepsilon_{Ab} . E^{inc} d^3 r'$ is the electric field scattered from the background structure with $G_{Ab}^E(r,r',\omega)$ being its electric dyadic Green's function. $E_{Ar}^{sca}$ is the residual scattered electric field, i.e., the difference between the electric fields scattered from resonator A and its background structure. By taking $E_{Ar}^{sca}\big|_{r \in \{\varepsilon_A \neq \varepsilon_{med}\}} = \sum_m \alpha_m \tilde{E}_{Am}$, like Eq. (3), and then using the procedure Eq. (6) was derived, we can obtain:

$$\mathbf{H}_A(\omega) \begin{bmatrix} j(\omega - \tilde{\omega}_1)\alpha_1 \\ j(\omega - \tilde{\omega}_2)\alpha_2 \\ \vdots \end{bmatrix} = -\frac{j\omega}{4} \iiint_{\varepsilon_A \neq \varepsilon_{Ab}} (\varepsilon_A - \varepsilon_{Ab}) T_{Ab}(E^{inc}) . \begin{bmatrix} \tilde{E}_{A1}^* \\ \tilde{E}_{A2}^* \\ \vdots \end{bmatrix} d^3 r, \tag{13}$$

where $T_{Ab}(E^{inc}) = E^{inc} + E_{Ab}^{sca} = E^{inc} + \omega^2 \mu_0 \iiint_{\varepsilon_{Ab} \neq \varepsilon_{med}} G_{Ab}^E . \Delta \varepsilon_{Ab} . E^{inc} d^3 r'$ is the total electric field that the sources create when the background structure replaces the resonator in the medium. Eq. (13) is a generalization of Eq. (6), since we can deduce Eq. (6) from Eq. (13) by assuming a free-space background ($\varepsilon_{Ab} = \varepsilon_{med}$). However, unlike Eq. (6), the non-principal modes have a negligible coupling to the incident wave in Eq. (13) as, by definition, their electric fields ($\tilde{E}_{An}^*$) are small in the region with $\varepsilon_A \neq \varepsilon_{Ab}$. Hence, if the principal and non-principal modes of the resonator are orthogonal, we can neglect all the elements related to the non-principal modes in $\mathbf{H_A}$ and $|\alpha\rangle$.

We can then express the scattered fields and the absorption of resonator A as a superposition of those from the background structure and the principal modes [35]:

$$E_A^{sca} \approx E_{Ab}^{sca} + \sum_m \alpha_m \tilde{E}_{Am}^{sca}, \tag{14}$$

$$A_A \approx A_{Ab} - 2\omega \text{Im}\left[\langle\alpha|\mathbf{H_A}|\alpha\rangle\right], \tag{15}$$

Where $A_{Ab}$ is the absorption of the background structure and same as Eq. (11), Eq. (15) can be further simplified to $A_A = A_{Ab} + \langle\alpha|2\mathbf{\Gamma_A^{nr}}|\alpha\rangle$ when the resonator modes are orthogonal and spectrally narrow. The approximations in Eqs. (14) and (15) are valid only if the background structure is appropriately chosen, i.e., either or both of the following conditions are satisfied [35]: (1) the region with $\varepsilon_A \neq \varepsilon_{Ab}$ is much smaller than the region with $\varepsilon_A \neq \varepsilon_{med}$; (2) the total electric field created by the background structure is small ( $T_{Ab}(E^{inc}) \ll E_{Ar}^{sca}$ ) in the region with $\varepsilon_A \neq \varepsilon_{Ab}$. The background terms in Eqs. (14) and (15) capture both the non-resonant terms and the collective responses from the resonator's non-principal modes in Eqs. (10) and (11). The background terms can be acquired from either of analytical calculations, numerical simulations, or experimental measurements on the background structure.

## 2.3. Resonator ports and matrix formulations

We assume that the electromagnetic fields in the medium can be decomposed into a linear superposition of some orthogonal waveguide modes. As a scatterer, resonator A disrupts this orthogonality relation inducing the coupling of the different waveguide modes to each other. Yet, the orthogonality relation still holds outside the resonator (regions with $\varepsilon_A = \varepsilon_{med}$), and thus the waveguide modes can expand the fields outside the resonator. Each waveguide mode coupled to the resonator (delivering the incident energy) corresponds to a resonator port. In this regard, the number of the resonator ports ($\mathbf{N}$) is the same as the number of the waveguide modes coupled to it. Although the number of ports may be infinite, we practically only consider a few of them by neglecting the ones which carry a relatively small amount of energy. In general, the waveguide modes may be of any type, e.g., spherical waves, plane waves, TEM modes over a coaxial cable, etc. However, in this paper, we only consider the waveguide modes with planar phase fronts, common in photonics [7].

Suppose that our medium is a kind of waveguide structure, with the $z$ axis being the waveguide axis. The region outside of resonator A can be divided into the two regions A1 and A2 with the reference planes $z = z_{A1}$ and $z = z_{A2}$, respectively (Figure 1(a)). We take $\beta_n$, $E_n^w(x,y)$ and $H_n^w(x,y)$ as the propagation constant and the normalized electric and magnetic field patterns (over $z = 0$) of the $n^{th}$ waveguide mode, respectively. $\beta_n$ is positive for forward-propagating modes and negative for backward-propagating modes. Then, the orthonormality relation between the waveguide modes can be written as [3]:

$$U_{ij} = \frac{1}{4}\iint_{S_w}\left[\left(E_i^w\right)^* \times H_j^w + E_j^w \times \left(H_i^w\right)^*\right]\cdot d\mathbf{s} = \pm\delta_{ij}, \tag{16}$$

where $S_w$ is a cross-sectional plane of the waveguide perpendicular to $z$ axis and $\delta_{ij}$ denotes the Kronecker delta with the plus and minus sign being for the forward and backward propagating modes, respectively. Without loss of generality, we index the waveguide modes such that $\beta_n > 0$ for $1 \leq n \leq N/2$ and $\beta_n = -\beta_{n-N/2}$ for $N/2+1 \leq n \leq N$ as illustrated in Figure 1(a).

We can expand the total electric field propagating towards and away from resonator A as $\sum_n S_{q_n,n}^{+A} E_n^w e^{-j\beta_n(z-z_{q_n})}$ and $\sum_n S_{q_n,n}^{-A} E_n^w e^{j\beta_n(z-z_{q_n})}$ with $n$ being the port number and $q_n$ being the reference plane of the port: $q_n = A1, A2$ if $\beta_n > 0$ and $\beta_n < 0$, respectively. "+" and "−" signs denote propagation toward and away from the resonator, respectively. Due to the normalization condition in Eq. (16), the unitless components $S_{q_n,n}^{\pm A}$ are the power waves [4], with $\left|S_{q_n,n}^{+A}\right|^2$ and $\left|S_{q_n,n}^{-A}\right|^2$ corresponding to the normalized power carried by the $n^{th}$ waveguide mode toward and away from resonator A, respectively. Without loss of generality, we assign the same phase angle to all the normalized electric field patterns. $S_{q_n,n}^{+A}$ terms can be written in vector form as $\left|S^{+A}\right\rangle = \left[S_{q_n,n}^{+A}\right]_{N\times 1} = \begin{bmatrix}\left|S_{A1}^{+A}\right\rangle \\ \left|S_{A2}^{+A}\right\rangle\end{bmatrix}$, with $\left|S_{A1}^{+A}\right\rangle$ and $\left|S_{A2}^{+A}\right\rangle$ being the input power wave vectors for the ports in regions A1 and A2, respectively. Similarly, $S_{q_n,n}^{-A}$ terms can be written in vector form as $\left|S^{-A}\right\rangle = \left[S_{q_n,n}^{-A}\right]_{N\times 1} = \begin{bmatrix}\left|S_{A1}^{-A}\right\rangle \\ \left|S_{A2}^{-A}\right\rangle\end{bmatrix}$ with $\left|S_{A1}^{-A}\right\rangle$ and $\left|S_{A2}^{-A}\right\rangle$ being the output power wave vectors for the ports in regions A1 and A2, respectively.

By assuming that the incident field does not have any non-propagating waveguide components, we can expand the incident electric field as $E^{inc}(r) = \sum_n S_{q_n,n}^{+A} E_n^w(x,y) e^{-j\beta_n(z-z_{q_n})}$. Then, via inserting this expansion into Eq. (13), we get the following matrix form for the CM equation:

$$j\omega|\alpha\rangle = j\tilde{\boldsymbol{\omega}}_A|\alpha\rangle + \mathbf{K}_{A1}^{A(0)}\left|S_{A1}^{+A}\right\rangle + \mathbf{K}_{A2}^{A(0)}\left|S_{A2}^{+A}\right\rangle = \mathbf{P}_A^{(0)}|\alpha\rangle + \mathbf{K}^{A(0)}\left|S^{+A}\right\rangle, \tag{17}$$

where $\tilde{\boldsymbol{\omega}}_A = \boldsymbol{\Omega}_A + j\boldsymbol{\Gamma}_A$ with $\boldsymbol{\Omega}_A$ and $\boldsymbol{\Gamma}_A$ being diagonal $M \times M$ matrices with the elements $\Omega_{Am}$ and $\Gamma_{Am}$, respectively. $\mathbf{P}_A^{(0)} = j\tilde{\boldsymbol{\omega}}_A$ is defined as the pole matrix of resonator A. We have defined the input-coupling matrix $\mathbf{K}^{A(0)}$ as:

$$\mathbf{K}^{A(0)} = \left[\mathbf{K}_{A1}^{A(0)}\ \mathbf{K}_{A2}^{A(0)}\right] = \mathbf{H}_A^{-1}\left[v_{mn} = \left(-\frac{j\omega}{4}\right)\iiint_{\varepsilon_A \neq \varepsilon_{Ab}}(\varepsilon_A - \varepsilon_{Ab}).T_{Ab}\left(E_n^w e^{-j\beta_n(z-z_{q_n})}\right).\tilde{E}_{Am}^* d^3r\right]_{M\times N}. \tag{18}$$

The elements of both the pole and input-coupling matrices have the unit of inverse time. Furthermore, the notation *(0)* in the superscripts indicates that they are obtained when resonator A is the only scatterer in the medium.

Just like a dipole antenna [6], the electric field scattered from resonator A (Eq. (14)) can be decomposed into propagating and non-propagating components as $E_A^{sca} = E_A^{sca,p} + E_A^{sca,np}$. The propagating component $E_A^{sca,p}$ carries the power radiated away from the resonator, and from Eq. (14), can be further decomposed as $E_A^{sca,p} = E_{Ab}^{sca,p} + \sum_m \alpha_m \tilde{E}_{Am}^{sca,p}$. The non-propagating component $E_A^{sca,np}$ corresponds to the energy stored in the resonator near-field [6]. In the CMT framework, it is more usual to express the total fields radiated away from the resonator rather than expressing the scattered fields [22]. By expanding $E^{inc} + E_A^{sca,p}$ as a superposition of the waveguide modes in each region and then, applying the conjugated Lorentz reciprocity theorem (similar procedure as the CMT for waveguides [3, 4]), we find the formula for the output power wave vector of the resonator [35]:

$$\left|S^{-A}\right> \approx \mathbf{S}^{Ab}\left|S^{+A}\right> + \mathbf{F}^{A(0)}\left|\alpha\right>. \tag{19}$$

Here, $\mathbf{S}^{Ab} = \begin{bmatrix} S_{11}^{Ab} & S_{12}^{Ab} \\ S_{21}^{Ab} & S_{22}^{Ab} \end{bmatrix}$ is the scattering matrix [4] of the background structure, which can be acquired from either analytical calculations, numerical simulations, or experimental measurements. $\mathbf{F}^{A(0)}$ is the output-coupling matrix of the resonator modes:

$$\mathbf{F}^{A(0)} = \begin{bmatrix} \mathbf{F}_{A1}^{A(0)} \\ \mathbf{F}_{A2}^{A(0)} \end{bmatrix} = \mathbf{S}^{Af} \left[ v_{nm} = \left(-\frac{j\omega}{4}\right) \iiint_{\varepsilon_A \neq \varepsilon_{med}} \Delta\varepsilon_A . \tilde{E}_{Am} . \left(E_n^w\right)^* e^{j\beta_n(z-z_{q_n})} d^3r \right]_{N \times M}, \tag{20}$$

where $\mathbf{S}^{Af} = \begin{bmatrix} 0 & \mathbf{D}(z_{A2} - z_{A1}) \\ \mathbf{D}(z_{A2} - z_{A1}) & 0 \end{bmatrix}$ is the scattering matrix for the free space slab between the reference planes with the delay matrix $\mathbf{D}(z_{A2} - z_{A1})$ defined as the $N/2 \times N/2$ diagonal matrix of the elements $d_{nn} = e^{-j\beta_n(z_{A2}-z_{A1})}$ with $1 \leq n \leq N/2$. The elements in $\mathbf{F}^{A(0)}$ are unitless. The approximation in Eq. (19) is valid only if the background structure is appropriately chosen.

By considering $A_{Ab} = \left<S^{+A}|S^{+A}\right> - \left<S^{+A}|\left(\mathbf{S}^{Ab}\right)^\dagger \mathbf{S}^{Ab}|S^{+A}\right>$, we can also rewrite the formula for the resonator absorption (Eq. (15)) in the matrix form:

$$A_A = \left<S^{+A}|S^{+A}\right> - \left<S^{+A}|\left(\mathbf{S}^{Ab}\right)^\dagger \mathbf{S}^{Ab}|S^{+A}\right> - 2\omega \operatorname{Im}\left[\left<\alpha|\mathbf{H}_A|\alpha\right>\right], \tag{21}$$

Moreover, by writing the energy balance relations for Eqs. (17) and (19), and then, combining the results with Eq. (21), we could get [35]:

$$\left(\mathbf{F}^{A(0)}\right)^{\dagger}\mathbf{F}^{A(0)} = -\left[\left(\mathbf{P}_{A}^{(0)}\right)^{\dagger}\left(\mathbf{H}_{A}\right)^{\dagger} + \mathbf{H}_{A}\mathbf{P}_{A}^{(0)}\right] \approx 2\mathbf{\Gamma}_{A}^{r}, \tag{22}$$

$$\mathbf{\kappa}^{A(0)} = -\mathbf{H}_{A}^{-1}\left(\mathbf{F}^{A(0)}\right)^{\dagger}\mathbf{S}^{Ab}. \tag{23}$$

The approximation in Eq. (22) is valid when the resonator modes are orthogonal, and the materials are low-dispersion and low-loss. In this scenario, like $\mathbf{\Gamma}_{A}^{r}$, $\left(\mathbf{F}^{A(0)}\right)^{\dagger}\mathbf{F}^{A(0)}$ becomes diagonal and we get $\sum_{n}\left|f_{nm}^{A(0)}\right|^{2} = 2\Gamma_{Am}^{r}$, implying that the radiative decay rate for the $m^{th}$ mode can be broken into the partial contributions from different ports as $\Gamma_{Am}^{r} = \sum_{n}\Gamma_{Am}^{r,n}$ with $\Gamma_{Am}^{r,n} = \left|f_{nm}^{A(0)}\right|^{2}/2$. In this regard, we can decompose the matrix $\mathbf{\Gamma}_{A}$ as $\mathbf{\Gamma}_{A} = \mathbf{\Gamma}_{A}^{nr} + \mathbf{\Gamma}_{A}^{r} = \mathbf{\Gamma}_{A}^{nr} + \sum_{n}\mathbf{\Gamma}_{A}^{r,n}$.

If the resonator's background structure is made of low-loss materials, we can take $\mathbf{S}^{Ab}$ to be unitary [4] to get $\left(\mathbf{S}^{Ab}\right)^{-1} = \left(\mathbf{S}^{Ab}\right)^{\dagger}$. Furthermore, from Eq. (9), we can approximate that $\mathbf{H}_{A} \approx \mathbf{I}$ when the resonator modes are orthogonal. Combining these approximations with Eqs. (22) and (23) leads to' $\mathbf{\kappa}^{A(0)}\left(\mathbf{\kappa}^{A(0)}\right)^{\dagger} = 2\mathbf{\Gamma}_{A}^{r}$ consistent with the conventional TCMT models [22, 24, 25, 27].

### 2.4. Generalization of the mode basis set

So far, we have used the resonator's QNMs as the mode basis set to expand the scattered fields inside the resonator. However, it is sometimes easier to use the QNMs of a different resonator with similar but simpler geometry as the mode basis set. As depicted in Figure 1(c), we consider a resonator A' with the dielectric function $\varepsilon_{A'} = \varepsilon_{A} + \delta\varepsilon$ where $\varepsilon_{A}$ is the dielectric function for the resonator we initially considered in Figure 1(a). $\delta\varepsilon$ is the difference between the dielectric functions of the two resonators, which is assumed to be relatively small, so the QNMs of resonator A can approximately expand the scattered fields in resonator A' as well. One can therefore write the Hamiltonian, pole and input-coupling matrices for resonator A' approximately as [35]:

$$\mathbf{H}_{A'} = \mathbf{H}_{A} + \delta\mathbf{H}, \tag{24}$$

$$\mathbf{P}_{A'}^{(0)} = \mathbf{P}_{A}^{(0)} + \delta\mathbf{P}, \tag{25}$$

$$\mathbf{\kappa}^{A'(0)} = \mathbf{\kappa}^{A(0)} + \delta\mathbf{\kappa}, \tag{26}$$

with $\delta\mathbf{H}$, $\delta\mathbf{P}$ and $\delta\mathbf{\kappa}$ defined as:

$$\delta\mathbf{H} = \left[v_{mn} = \frac{1}{4}\iiint_{\delta\varepsilon \neq 0} \delta\varepsilon \tilde{E}_{An}.\tilde{E}_{Am}^{*} d^{3}r\right]_{M \times M} \tag{27}$$

$$\delta \mathbf{P} = -\mathbf{H}_{A'}^{-1}\delta\mathbf{H}\mathbf{P}_{A}^{(0)}, \tag{28}$$

$$\delta \mathbf{\kappa} = \mathbf{H}_{A'}^{-1}\left[v_{mn} = \left(-\frac{j\omega}{4}\right)\iiint_{\delta\varepsilon \neq 0}\delta\varepsilon.T_{Ab}\left(E_{n}^{w}e^{-j\beta_{n}\left(z-z_{q_{n}}\right)}\right).\tilde{E}_{Am}^{*}d^{3}r\right]_{M\times N} - \mathbf{H}_{A'}^{-1}\delta\mathbf{H}\mathbf{\kappa}^{A(0)}. \tag{29}$$

Equations (25) and (28) reveal that the pole matrix $\mathbf{P}_{A'}^{(0)}$ is diagonal only if we use the resonator's own QNMs as the basis set. Non-diagonality of $\mathbf{P}_{A'}^{(0)}$ is indicative of coupling between the resonator modes. Eqs. (14) and (15) can still be used to obtain the output power wave vector and the absorption for resonator A'.

### 2.5. Signal flow graph, characteristic equation, and scattering matrix

In Control engineering, Eqs. (17) and (19) constitute a set of linear state-space equations [32] with $|\alpha\rangle$ as the state vector that characterizes the system (resonator A) state and $|S^{-A}\rangle$ as the output vector. A more intuitive way to express the state-space equations is to graphically represent the dynamics they describe using signal flow graphs (SFG) [4, 32]. To do so, we first rewrite Eq. (17) as:

$$|\dot{\alpha}\rangle = \mathbf{P}_{A}^{(0)}|\alpha\rangle + \mathbf{\kappa}^{A(0)}|S^{+A}\rangle, \tag{30a}$$

$$|\alpha\rangle = \frac{1}{j\omega}|\dot{\alpha}\rangle. \tag{30b}$$

Then, by considering $|\alpha\rangle$ and $|\dot{\alpha}\rangle$ as the nodes, we get the SFG shown in Figure 1(d). Here, the $\mathbf{P}_{A}^{(0)}$ arrow is the feedback, and the loop created by the $\mathbf{P}_{A}^{(0)}$ and the $\frac{1}{j\omega}$ arrows represents the resonant energy circulation inside the resonator. SFGs are very useful for representing and interpreting the electromagnetic interaction dynamics that the CM equations describe. They are also powerful tools to solve the CM equations using Mason's theorem [32].

From Eq. (17), we can calculate the amplitude vector of resonator A as follows:

$$|\alpha\rangle = \left(j\omega\mathbf{I} - \mathbf{P}_{A}^{(0)}\right)^{-1}\mathbf{\kappa}^{A(0)}|S^{+A}\rangle. \tag{31}$$

Equation (31) reveals that we can acquire the resonator poles using the eigenvalue equation $\left|s\mathbf{I} - \mathbf{P}_{A}^{(0)}\right| = 0$, called the *characteristic equation* [32], where $s = j\omega$. Solving the characteristic equation for $\omega$ gives the complex resonance frequencies of the resonator. When the pole matrix is diagonal, each mode (e.g., the $m^{th}$ mode) only has a single resonance frequency $\tilde{\omega}_{Am} = \Omega_{Am} + j\Gamma_{Am}$. This result is consistent with the fact that each of the resonator QNMs is associated with a single resonance frequency, and again implies that the pole matrix becomes diagonal only when the resonator's own QNMs are chosen as the basis set, as we discussed in the previous subsection. In contrast, when the basis set is not the resonator's own QNMs, the pole matrix becomes non-diagonal, making the resonance frequencies of different modes inseparable.

By plugging Eq. (31) into Eqs. (19) and (21), we can calculate the output power wave vector and the absorption of the resonator. Moreover, by combining Eq. (31) with Eqs. (19) and (23), we can obtain the following expression for the scattering matrix of the resonator:

$$\mathbf{S}^{\mathbf{A}} = \left[ \mathbf{I} - \mathbf{F}^{\mathbf{A(0)}} \left( j\omega \mathbf{I} - \mathbf{P}_{\mathbf{A}}^{(0)} \right)^{-1} \mathbf{H}_{\mathbf{A}}^{-1} \left( \mathbf{F}^{\mathbf{A(0)}} \right)^{\dagger} \right] \mathbf{S}^{\mathbf{Ab}} . \tag{32}$$

An alternative way to obtain Eqs. (31) and (32) is applying Mason's theorem [32] to the SFG in Figure 1(d).

## 2.6. Parameter retrieval by reverse engineering

The application of Eqs. (17), (19) and (21) requires the information of the complex resonance frequencies as well as the input and output coupling coefficients for the resonator modes as a pre-requisite. To obtain the complex resonance frequencies, one must solve the Maxwell's equations with the resonator boundary conditions [19]. Moreover, to obtain the input and output coupling coefficients, one must calculate the overlap integrals in Eqs. (18) and (20) from the field distributions for both the resonator and the waveguide modes. Whether we acquire the complex resonance frequencies and the coupling coefficients by analytical calculations or numerical simulations, such a theoretical approach is very demanding, which hinders the versatile application of the developed CMT model. Furthermore, it is challenging to get the three-dimensional field profiles and then to calculate the overlap integrals in the experimental situations. To alleviate these difficulties, we introduce the kappa-estimation and F-estimation lemmas to reverse-engineer the decay rates and the coupling coefficients from the far-field data without calculating complicated integrals. The detailed derivation process of these lemmas can be found in SI section 1.5 [35].

**Kappa-estimation lemma.** Suppose that: 1) Resonator A is made of low-dispersion and lossy ($\mathrm{Im}[\varepsilon_A] \neq 0$) materials, 2) Resonator A's QNMs are orthogonal and used as the mode basis set, and 3) the background structure is lossless. Moreover, assume that the resonator absorption spectra around $\omega = \Omega_{Am}$ is known when it is excited from different ports. The absorption spectra around $\omega = \Omega_{Am}$ typically have a Lorentzian-like lineshape superimposed over a non-Lorentzian background. For each port (e.g., $n^{th}$ port), we acquire the Lorentzian lineshape by the background-subtraction (Figure 1(e)), from which we can determine the resonance frequency $\Omega_{Am}$, the full width at half maximum (FWHM) $\Delta\omega_{Am}$, and the maximum normalized absorption ($0 < A_{mn}^{\max} \leq 1$) of the $m^{th}$ mode. It has been found that $\Omega_{Am}$ and $\Delta\omega_{Am}$ are almost the same for excitations from different ports [35]. Then, we can estimate the nonradiative decay rate and the input-coupling coefficients of the $m^{th}$ mode as follows:

$$\Gamma_{Am}^{nr} = \frac{\Delta\omega_{Am}}{4}(1 \pm \Lambda), \tag{33}$$

$$\left| \kappa_{mn}^{A(0)} \right|^2 = \frac{\Delta\omega_{Am}}{2} \frac{A_{mn}^{\max}}{1 \pm \Lambda}, \tag{34}$$

with $\Lambda = \sqrt{1 - \sum_n A_{mn}^{\max}}$. Eqs. (33) and (34) give us two solution sets. Moreover, the lemma does not directly give us the phase angles of the input-coupling coefficients. We can determine the valid solution set and the phase angles by fitting the scattering spectra. An alternative way is to look at the dominant multipoles and the radiation pattern of the mode. Once the mode input-coupling coefficients are found, its output coupling coefficients can be obtained from Eq. (23).

**F-estimation lemma.** Suppose that resonator A's QNMs are orthogonal and used as the mode basis set. Also, we assume that the FWHM ($\Delta\omega_{Am}$) for the $m^{th}$ mode is known. We must first find the scattering matrix ($\mathbf{S^A}$) of the resonator at the frequency $\omega = \Omega_{Am}$ by exciting the resonator from each port and measuring the fields scattered into the other ports using numerical simulations or experimental measurements (Figure 1(f)). Then, we can estimate the output-coupling coefficients and nonradiative decay rate of the $m^{th}$ mode as follows [35]:

$$\begin{bmatrix} \left|f_{1m}^{A(0)}\right|^2 & \cdots & f_{1m}^{A(0)}\left(f_{Nm}^{A(0)}\right)^* \\ \vdots & \ddots & \vdots \\ f_{Nm}^{A(0)}\left(f_{1m}^{A(0)}\right)^* & \cdots & \left|f_{Nm}^{A(0)}\right|^2 \end{bmatrix} \approx \frac{\Delta\omega_{Am}}{2}\left[\mathbf{I} - \mathbf{S^A}\left(\mathbf{S^{Ab}}\right)^{-1}\right]\bigg|_{\omega=\Omega_{Am}}, \qquad (35)$$

$$\Gamma_{Am}^{nr} \approx \frac{\Delta\omega_{Am}}{2} - \sum_n \frac{\left|f_{nm}^{A(0)}\right|^2}{2} = \frac{\Delta\omega_{Am}}{2}\left[1 - \frac{1}{2}\text{Tr}\left(\mathbf{I} - \mathbf{S^A}\left(\mathbf{S^{Ab}}\right)^{-1}\right)\right]\bigg|_{\omega=\Omega_{Am}}. \qquad (36)$$

In Eq. (35), real parts of the diagonal elements of the matrix $\frac{\Delta\omega_{Am}}{2}\left[\mathbf{I} - \mathbf{S^A}\left(\mathbf{S^{Ab}}\right)^{-1}\right]\bigg|_{\omega=\Omega_{Am}}$ approximately give us the magnitudes for the output coupling coefficients of the $m^{th}$ mode. We can estimate the phase angles of the output coupling coefficients using the methods introduced to achieve the same purpose for the Kappa-estimation lemma. After determining the mode output-coupling coefficients, its input-coupling coefficients can be estimated from Eq. (23). Unlike the Kappa-estimation lemma, the F-estimation lemma gives a single solution and does not require the resonator to be lossy and the background structure to be lossless. However, this improved versatility is compromised by a higher computational cost as we should first acquire the resonator scattering matrix to use this lemma.

In summary, based on Maxwell's equations and QNM theory, this section established the BBCMT framework for a single resonator, with Eqs. (17), (19), (21), (22), and (23) constituting the main equations. Notably, the framework allows expressing a resonator's responses as perturbations to those of a user-defined background structure that captures the resonator's non-resonant responses and non-principal modes' responses. The background structure is like a black-box, for which the framework merely uses the input-output relationship without involving the transformation dynamical details. There is no unique correct choice for the background structure; The user can determine it according to their principal areas of interest in the resonator and the desired level of accuracy and complexity. In this regard, BBCMT is a malleable and arbitrarily-analytical platform, such that it can be highly accurate or only give a rough estimation of the resonator behavior, depending on the considered approximations and background structure. Besides developing the BBCMT equations, we introduced the Kappa-estimation and F-estimation lemmas to reverse-engineer the BBCMT models' parameters from the calculated, simulated, or measured far-field data. Although we assumed non-magnetic and isotropic materials to derive this section's formulations for simplicity, the approach is similar for the more general case without these constraints.

## 3. BBCMT formulations for coupled resonators

In this section, we give a brief derivation of the BBCMT formulas for a system of coupled resonators and discuss their potential impacts. In subsection 3.1, we derive the CM equations for a system of two coupled resonators. Then, we derive the equations for the stored energy, scattering, and absorption for the system consisting of the two coupled resonators in subsection 3.2. Finally, in subsection 3.3, we discuss the possible simplifying approximations for the special case when the resonators' background structures are non-resonant in the operating frequency range.

### 3.1. Loading effects and coupled-mode equations

Now, we consider a coupled-resonator system with a resonator B added to the configuration in Figure 1(a), as shown in Figure 2(a). The fields scattered from the coupled system can be decomposed into a summation of the fields scattered from each of the subsystem resonators as:

$$E^{sca} = E_A^{sca} + E_B^{sca}, \tag{37a}$$

$$H^{sca} = H_A^{sca} + H_B^{sca}, \tag{37b}$$

where $E_A^{sca}$ and $H_A^{sca}$ satisfy:

$$\nabla \times E_A^{sca} = -j\omega\mu_0 H_A^{sca}, \tag{38a}$$

$$\nabla \times H_A^{sca} = j\omega\varepsilon_A E_A^{sca} + j\omega\Delta\varepsilon_A \left(E^{inc} + E_B^{sca}\right), \tag{38b}$$

and, $E_B^{sca}$ and $H_B^{sca}$ satisfy:

$$\nabla \times E_B^{sca} = -j\omega\mu_0 H_B^{sca}, \tag{39a}$$

$$\nabla \times H_B^{sca} = j\omega\varepsilon_B E_B^{sca} + j\omega\Delta\varepsilon_B \left(E^{inc} + E_A^{sca}\right), \tag{39b}$$

where $\varepsilon_B$ is the dielectric function when only resonator B is present and $\Delta\varepsilon_B = \varepsilon_B - \varepsilon_{med}$. Eqs. (38) and (39) resemble Eqs. (2) except that each resonator is now excited by the electric field scattered from the other resonator beside the incident field. Hence, the fields scattered from the two resonators are not independent of each other (Figure 2(b)), indicating that each resonator acts as a load for the other one just as two circuit networks connected in cascade [39]. This mutual loading effect has two major consequences on the modes inside each resonator: modification of the complex resonance frequencies and modulation of the coupling to the incident fields. Modification of the complex resonance frequencies is a result of self-induced back-action [5] of the resonators' modes due to back-and-forth scattering of fields between the resonators (Figure 2(b)). From the energy flow perspective, modification of the complex resonance frequencies occurs as back-and-forth scattering of fields between the resonators creates new routes for their modes' energy circulation alongside the internal energy circulation routes. As the second consequence, modulation of the coupling to the incident fields occurs because, besides the direct-coupling channel to the incident field, each resonator now has an indirect-coupling channel through the scattered fields from the other resonator (Figure 2(b)).

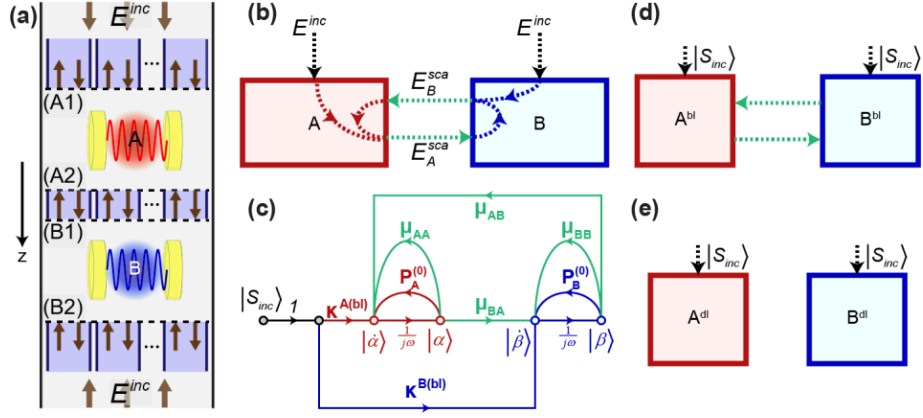

**Figure 2.** BBCMT model for the system of coupled resonators. (a) System of the coupled resonators A and B in the medium. (b) Diagram of energy circulation for the system of coupled resonators reveals the mutual loading effect between them. (c) The SFG for the system of the coupled resonators. Output-coupling of the system modes is not shown in the SFG for simplicity. (d, e) The equivalent scenarios of the coupled background-loaded resonators (d) and the decoupled-loaded resonators (e).

The mutual loading effect can be divided into non-resonant and resonant loading effects due to non-resonant and resonant scattering of the fields from the resonators, respectively. Moreover, all the modes from both resonators are involved in the resonant loading effect while we are interested in only using the principal modes in our modeling. Hence, we again resort to the background perturbation method. We show the dielectric functions of the background structures for resonators A and B by $\varepsilon_{Ab}$ and $\varepsilon_{Bb}$, respectively. By following the procedure used to derive Eq. (14) together with Eqs. (38) and (39), and under the condition that the background structures are chosen properly (see section 2.2), we can obtain the following equations for the electric fields scattered from the resonators A and B:

$$E_A^{sca} \approx E_{Ab}^{sca(0)} + \sum_m \alpha_m \tilde{E}_{Am}^{sca(0)} + \omega^2 \mu_0 \iiint_{\varepsilon_{Ab} \neq \varepsilon_{med}} G_{med}^E \cdot \Delta\varepsilon_{Ab} \cdot T_{Ab}\left(E_B^{sca}\right) d^3 r', \tag{40a}$$

$$E_B^{sca} \approx E_{Bb}^{sca(0)} + \sum_k \beta_k \tilde{E}_{Bk}^{sca(0)} + \omega^2 \mu_0 \iiint_{\varepsilon_{Bb} \neq \varepsilon_{med}} G_{med}^E \cdot \Delta\varepsilon_{Bb} \cdot T_{Bb}\left(E_A^{sca}\right) d^3 r', \tag{40b}$$

where $\Delta\varepsilon_{Bb} = \varepsilon_{Bb} - \varepsilon_{med}$ and $E_{Ab}^{sca(0)}$, $\tilde{E}_{Am}^{sca(0)}$, $E_{Bb}^{sca(0)}$ and $\tilde{E}_{Bk}^{sca(0)}$ are defined as the similar terms in Eq. (14). We have added the notation *(0)* in the superscripts of the electric field terms for resonator A (B) to indicate their belonging to the situation where resonator A (B) is solely present in the medium. The integral term on the right-hand-side of Eq. (40a) ((40b)) represents the scattering of $E_B^{sca}$ ($E_A^{sca}$) from the background of resonator A (B). As a result of these integral terms, Eqs. (40a) and (40b) are not independent of each other. Nevertheless, by combining Eqs. (40a) and (40b) together, we can decouple them to obtain the following equations for $E_A^{sca}$ and $E_B^{sca}$:

$$E_A^{sca} = E_{Ab}^{sca} + \sum_m \alpha_m \tilde{E}_{Am}^{sca(e)} + \sum_k \beta_k \tilde{E}_{Bk}^{sca(o)}, \tag{41a}$$

$$E_B^{sca} = E_{Bb}^{sca} + \sum_k \beta_k \tilde{E}_{Bk}^{sca(e)} + \sum_m \alpha_m \tilde{E}_{Am}^{sca(o)}, \tag{41b}$$

where $E_{Ab}^{sca} = E_{Ab}^{sca(e)} + E_{Bb}^{sca(o)}$ and $E_{Bb}^{sca} = E_{Bb}^{sca(e)} + E_{Ab}^{sca(o)}$ with $E_{Ab}^{sca(e)} = E_{Ab}^{sca(0)} + \sum_{j=2i} E_{Ab}^{sca(j)}$,

$E_{Ab}^{sca(o)} = \sum_{j=2i+1} E_{Ab}^{sca(j)}$, $E_{Ab}^{sca(2i)} = \omega^2 \mu_0 \iiint_{\varepsilon_{Ab} \neq \varepsilon_{med}} G_{med}^E \cdot \Delta \varepsilon_{Ab} \cdot T_{Ab}\left(E_{Ab}^{sca(2i-1)}\right) d^3 r'$ and

$E_{Ab}^{sca(2i+1)} = \omega^2 \mu_0 \iiint_{\varepsilon_{Bb} \neq \varepsilon_{med}} G_{med}^E \cdot \Delta \varepsilon_{Bb} \cdot T_{Bb}\left(E_{Ab}^{sca(2i)}\right) d^3 r'$. Also, $\tilde{E}_{Am}^{sca(e)} = \tilde{E}_{Am}^{sca(0)} + \sum_{j=2i} \tilde{E}_{Am}^{sca(j)}$ and

$\tilde{E}_{Am}^{sca(o)} = \sum_{j=2i+1} \tilde{E}_{Am}^{sca(j)}$ with $\tilde{E}_{Am}^{sca(2i)} = \omega^2 \mu_0 \iiint_{\varepsilon_{Ab} \neq \varepsilon_{med}} G_{med}^E \cdot \Delta \varepsilon_{Ab} \cdot T_{Ab}\left(\tilde{E}_{Am}^{sca(2i-1)}\right) d^3 r'$ and

$\tilde{E}_{Am}^{sca(2i+1)} = \omega^2 \mu_0 \iiint_{\varepsilon_{Bb} \neq \varepsilon_{med}} G_{med}^E \cdot \Delta \varepsilon_{Bb} \cdot T_{Bb}\left(\tilde{E}_{Am}^{sca(2i)}\right) d^3 r'$. Similar expressions hold for $E_{Bb}^{sca(e)}$, $E_{Bb}^{sca(o)}$, $\tilde{E}_{Bk}^{sca(e)}$ and $\tilde{E}_{Bk}^{sca(o)}$. $i$ is a positive integer in all the summations. $0^{th}$ order fields denote the direct scattering of the incident fields from each resonator. Higher-order scattering terms account for multiple back-and-forth scattering of the $0^{th}$ order fields between the two background structures. The background field terms in Eqs. (41) represent the fields scattered by each background structure when both resonators are replaced with their background structures, i.e., the system with $\varepsilon_b = \varepsilon_{med} + \Delta \varepsilon_{Ab} + \Delta \varepsilon_{Bb}$. From the above definitions for different scattering terms, we can deduce the following relations between them:

$$E_{Ab}^{sca} = E_{Ab}^{sca(0)} + \omega^2 \mu_0 \iiint_{\varepsilon_{Ab} \neq \varepsilon_{med}} G_{med}^E \cdot \Delta \varepsilon_{Ab} \cdot T_{Ab}\left(E_{Bb}^{sca}\right) d^3 r', \tag{42a}$$

$$E_{Bb}^{sca} = E_{Bb}^{sca(0)} + \omega^2 \mu_0 \iiint_{\varepsilon_{Bb} \neq \varepsilon_{med}} G_{med}^E \cdot \Delta \varepsilon_{Bb} \cdot T_{Bb}\left(E_{Ab}^{sca}\right) d^3 r', \tag{42b}$$

$$\tilde{E}_{Am}^{sca(e)} = \tilde{E}_{Am}^{sca(0)} + \omega^2 \mu_0 \iiint_{\varepsilon_{Ab} \neq \varepsilon_{med}} G_{med}^E \cdot \Delta \varepsilon_{Ab} \cdot T_{Ab}\left(\tilde{E}_{Am}^{sca(o)}\right) d^3 r', \tag{42c}$$

$$\tilde{E}_{Am}^{sca(o)} = \omega^2 \mu_0 \iiint_{\varepsilon_{Bb} \neq \varepsilon_{med}} G_{med}^E \cdot \Delta \varepsilon_{Bb} \cdot T_{Bb}\left(\tilde{E}_{Am}^{sca(e)}\right) d^3 r', \tag{42d}$$

$$\tilde{E}_{Bk}^{sca(o)} = \omega^2 \mu_0 \iiint_{\varepsilon_{Ab} \neq \varepsilon_{med}} G_{med}^E \cdot \Delta \varepsilon_{Ab} \cdot T_{Ab}\left(\tilde{E}_{Bk}^{sca(e)}\right) d^3 r', \tag{42e}$$

$$\tilde{E}_{Bk}^{sca(e)} = \tilde{E}_{Bk}^{sca(0)} + \omega^2 \mu_0 \iiint_{\varepsilon_{Bb} \neq \varepsilon_{med}} G_{med}^E \cdot \Delta \varepsilon_{Bb} \cdot T_{Bb}\left(\tilde{E}_{Bk}^{sca(o)}\right) d^3 r'. \tag{42f}$$

The scattered fields in Eqs. (41) can be either calculated in a step-wise manner, from their definitions, or by directly solving the integral equations (42). As we would see, for $E_{Ab}^{sca}$ and $E_{Bb}^{sca}$, it suffices to know their summation $E_b^{sca} = E_{Ab}^{sca} + E_{Bb}^{sca}$, which is the electric field scattered from the collective system of the background structures, i.e., the one corresponding to $\varepsilon_b = \varepsilon_{med} + \Delta \varepsilon_{Ab} + \Delta \varepsilon_{Bb}$. $E_b^{sca}$ can be acquired from numerical simulations or experimental measurements as well.

Now, we obtain the CM equations for the system of the coupled resonators. As explained, Eqs. (38) are like Eqs. (2) but with an additional source field. By plugging Eqs. (41) and the magnetic counterpart of Eq.

(41a) into Eqs. (38), and then following the procedure used to derive Eq. (6), we would obtain the CM equation for resonator A:

$$\mathbf{H_A}(\omega) \begin{bmatrix} j(\omega - \tilde{\omega}_{A1})\alpha_1 \\ j(\omega - \tilde{\omega}_{A2})\alpha_2 \\ \vdots \end{bmatrix} = -\frac{j\omega}{4} \iiint_{\varepsilon_A \neq \varepsilon_{Ab}} (\varepsilon_A - \varepsilon_{Ab}) T_{Ab}(E^{inc} + E_B^{sca}) \cdot \begin{bmatrix} \tilde{E}_{A1}^* \\ \tilde{E}_{A2}^* \\ \vdots \end{bmatrix} d^3r =$$

$$-\frac{j\omega}{4} \iiint_{\varepsilon_A \neq \varepsilon_{Ab}} (\varepsilon_A - \varepsilon_{Ab}) T_{Ab}(E^{inc} + E_{Bb}^{sca}) \cdot \begin{bmatrix} \tilde{E}_{A1}^* \\ \tilde{E}_{A2}^* \\ \vdots \end{bmatrix} d^3r - \frac{j\omega}{4} \iiint_{\varepsilon_A \neq \varepsilon_{Ab}} (\varepsilon_A - \varepsilon_{Ab}) T_{Ab}\left(\sum_m \alpha_m \tilde{E}_{Am}^{sca(o)}\right) \cdot \begin{bmatrix} \tilde{E}_{A1}^* \\ \tilde{E}_{A2}^* \\ \vdots \end{bmatrix} d^3r \quad (43)$$

$$-\frac{j\omega}{4} \iiint_{\varepsilon_A \neq \varepsilon_{Ab}} (\varepsilon_A - \varepsilon_{Ab}) T_{Ab}\left(\sum_k \beta_k \tilde{E}_{Bk}^{sca(e)}\right) \cdot \begin{bmatrix} \tilde{E}_{A1}^* \\ \tilde{E}_{A2}^* \\ \vdots \end{bmatrix} d^3r.$$

$T_{Ab}(E^{inc} + E_{Bb}^{sca})$ in Eq. (43) is in fact the total electric field created by the collective background structure since we can write $T_{Ab}(E^{inc} + E_{Bb}^{sca}) = E^{inc} + E_{Bb}^{sca} + E_{Ab}^{sca} = E^{inc} + E_b^{sca}$. Accordingly, we can rewrite this term as $T_b(E^{inc}) = E^{inc} + \omega^2 \mu_0 \iiint_{\varepsilon_b \neq \varepsilon_{med}} G_b^E \cdot \Delta\varepsilon_b \cdot E^{inc} d^3r'$ where $G_b^E$ is the electric dyadic Green's function for the collective background structure. We can expand the incident electric field term in Eq. (43) in terms of the medium waveguide modes as $E^{inc} = \sum_i S_{A1,i}^{+A} E_i^w e^{-j\beta_i(z-z_{A1})} + \sum_l S_{B2,l}^{+B} E_l^w e^{-j\beta_l(z-z_{B2})}$ with $\beta_i > 0$ and $\beta_l < 0$. By then inserting this expansion into Eq. (43), we can express it in matrix form:

$$j\omega|\alpha\rangle = \left(\mathbf{P_A^{(0)}} + \mathbf{\mu_{AA}}\right)|\alpha\rangle + \mathbf{\kappa^{A(bl)}}|S_{inc}\rangle + \mathbf{\mu_{AB}}|\beta\rangle, \quad (44)$$

where $\mathbf{P_A^{(0)}}$ is similar to that in Eq. (17) and $|S_{inc}\rangle = \begin{bmatrix} |S_{A1}^{+A}\rangle \\ |S_{B2}^{+B}\rangle \end{bmatrix}$ denotes the input power wave vector for the collective system of resonators. Moreover, $\mathbf{\kappa^{A(bl)}}$, $\mathbf{\mu_{AA}}$ and $\mathbf{\mu_{AB}}$ are defined as:

$$\mathbf{\kappa^{A(bl)}} = \begin{bmatrix} \mathbf{\kappa_{A1}^{A(bl)}} & \mathbf{\kappa_{B2}^{A(bl)}} \end{bmatrix} = \mathbf{H_A^{-1}} \left[ v_{mn} = \left(-\frac{j\omega}{4}\right) \iiint_{\varepsilon_A \neq \varepsilon_{Ab}} (\varepsilon_A - \varepsilon_{Ab}) \cdot T_b\left(E_n^w e^{-j\beta_n(z-z_{q_n})}\right) \cdot \tilde{E}_{Am}^* d^3r \right]_{M \times N}, \quad (45)$$

$$\mathbf{\mu_{AA}} = \mathbf{H_A^{-1}} \left[ x_{mn} = \left(-\frac{j\omega}{4}\right) \iiint_{\varepsilon_A \neq \varepsilon_{Ab}} (\varepsilon_A - \varepsilon_{Ab}) \cdot T_{Ab}\left(\tilde{E}_{An}^{sca(o)}\right) \cdot \tilde{E}_{Am}^* d^3r \right]_{M \times M}, \quad (46)$$

$$\mathbf{\mu_{AB}} = \mathbf{H_A^{-1}} \left[ x_{mk} = \left(-\frac{j\omega}{4}\right) \iiint_{\varepsilon_A \neq \varepsilon_{Ab}} (\varepsilon_A - \varepsilon_{Ab}) \cdot T_{Ab}\left(\tilde{E}_{Bk}^{sca(e)}\right) \cdot \tilde{E}_{Am}^* d^3r \right]_{M \times K}, \quad (47)$$

with **K** being the number of principal modes in resonator B and $q_n = A1, B2$ for $\beta_n > 0$ and $\beta_n < 0$, respectively. $\boldsymbol{\kappa}^{A(bl)}$ and $\boldsymbol{\mu}_{AA}$ denote the input-coupling matrix and the self-coupling matrix for resonator A when loaded with resonator B's background structure, respectively. These matrices capture the modification of the input-coupling and the complex resonance frequencies for resonator A due to non-resonant scattering from resonator B as well as the scattering from its non-principal modes. By decomposing $\tilde{E}_{An}^{sca(o)}$ in Eq. (46) into near-field (non-propagating) and radiative (propagating) components as $\tilde{E}_{An}^{sca(o)} = \tilde{E}_{An}^{sca(o),np} + \tilde{E}_{An}^{sca(o),p}$, we can further decompose the self-coupling matrix into near-field and radiative components as $\boldsymbol{\mu}_{AA} = \boldsymbol{\mu}_{AA}^{n} + \boldsymbol{\mu}_{AA}^{r}$. $\boldsymbol{\mu}_{AB}$ in Eq. (47) is the B-to-A coupling matrix that captures the energy transfer from the principal modes in resonator B to the principal modes in resonator A and, like $\boldsymbol{\mu}_{AA}$, can be broken into near-field and radiative components as $\boldsymbol{\mu}_{AB} = \boldsymbol{\mu}_{AB}^{n} + \boldsymbol{\mu}_{AB}^{r}$. Both $\boldsymbol{\mu}_{AA}$ and $\boldsymbol{\mu}_{AB}$ have the units of inverse time.

Parallel to resonator A, we can obtain the matrix format of the CM equation for resonator B as follows:

$$j\omega|\beta\rangle = \left(\mathbf{P}_B^{(0)} + \boldsymbol{\mu}_{BB}\right)|\beta\rangle + \boldsymbol{\kappa}^{B(bl)}|S_{inc}\rangle + \boldsymbol{\mu}_{BA}|\alpha\rangle, \tag{48}$$

where $\mathbf{P}_B^{(0)}$, $\boldsymbol{\mu}_{BB}$, $\boldsymbol{\kappa}^{B(bl)}$ and $\boldsymbol{\mu}_{BA}$ are defined similarly as Eqs. (45)-(47). Just as Eq. (13), we can eliminate all the elements related to the non-principal modes from Eqs. (44) and (48).

The SFG in Figure 2(c) expresses the electromagnetic interaction dynamics described by the CM Eqs. (44) and (48). One can see from the SFG that, in addition to the original feedback arrows $\mathbf{P}_A^{(0)}$ and $\mathbf{P}_B^{(0)}$, each resonator has an additional feedback branch: $\boldsymbol{\mu}_{AA}$ for resonator A and $\boldsymbol{\mu}_{BB}$ for resonator B. We define $A^{bl}$, the background-loaded resonator A, as the resonator with the pole matrix $\mathbf{P}_A^{(bl)} = \mathbf{P}_A^{(0)} + \boldsymbol{\mu}_{AA}$ and the input-coupling matrix $\boldsymbol{\kappa}^{A(bl)}$. This resonator is what we get when we keep resonator A but replace resonator B by its background structure and can be considered as a black-box that captures the loading effects on resonator A due to non-resonant scattering and resonant scattering of non-principal modes from resonator B. Unlike $\mathbf{P}_A^{(0)}$, the matrix $\mathbf{P}_A^{(bl)}$ is generally frequency-dependent, which causes each mode in resonator $A^{bl}$ to have multiple complex resonance frequencies rather than one. This result can be justified by noticing that $A^{bl}$ is, in fact, a Fabry-Perot-like resonator with resonator A and the background structure of resonator B as the end mirrors. Moreover, $\mathbf{P}_A^{(bl)}$ is generally non-diagonal, reflecting the couplings between the modes in the resonator $A^{bl}$. When the non-resonant scattering and the scattering from the non-principal modes of resonator B are small, we could simply consider the free space as its background structure. In such a scenario, Eqs. (42) and (46) give us $\boldsymbol{\mu}_{AA} = 0$ and thus, $\mathbf{P}_A^{(bl)} = \mathbf{P}_A^{(0)}$. Also, Eqs. (18) and (45) give us $\boldsymbol{\kappa}_{A1}^{A(bl)} = \boldsymbol{\kappa}_{A1}^{A(0)}$ and $\boldsymbol{\kappa}_{B2}^{A(bl)} = \boldsymbol{\kappa}_{A2}^{A(0)} \mathbf{D}(z_{B2} - z_{A2}) = \boldsymbol{\kappa}_{B2}^{A(0)}$, which merely describes a phase shift due to change of resonator A's second reference plane from A2 to B2. Accordingly, $A^{bl} \equiv A$, when the background structure of resonator B is free space. In parallel to $A^{bl}$, we can define $B^{bl}$, the background-loaded resonator B, as the resonator having the pole matrix $\mathbf{P}_B^{(bl)} = \mathbf{P}_B^{(0)} + \boldsymbol{\mu}_{BB}$ and the input-coupling matrix $\boldsymbol{\kappa}^{B(bl)}$. In terms of mode amplitudes and resonance frequencies, A and B coupled-resonator system is equivalent to system of the coupled background-loaded resonators, $A^{bl}$ and $B^{bl}$ (Figure 2(d)). Upon coupling, resonance frequencies of the resonators A and B become inseparable. We can find the resonance frequencies for the collective system of resonators by solving the collective characteristic equation of the system, which can be found from Eqs. (44) and (48):

$$\left|j\omega\mathbf{I}-\mathbf{P}^{(bl)}\right|=0, \tag{49}$$

where $\mathbf{P}^{(bl)}=\mathbf{P}^{(0)}+\boldsymbol{\mu}$ with $\mathbf{P}^{(0)}=\begin{bmatrix}\mathbf{P}_A^{(0)} & 0 \\ 0 & \mathbf{P}_B^{(0)}\end{bmatrix}$ and $\boldsymbol{\mu}=\begin{bmatrix}\boldsymbol{\mu}_{AA} & \boldsymbol{\mu}_{AB} \\ \boldsymbol{\mu}_{BA} & \boldsymbol{\mu}_{BB}\end{bmatrix}$.

As the SFG in Figure 2(c) manifests, the coupling terms $\boldsymbol{\mu}_{AB}$ and $\boldsymbol{\mu}_{BA}$ create further self-coupling channels for each resonator. For resonator A, this self-coupling channel includes $\boldsymbol{\mu}_{BA}$, $1/j\omega$ (from resonator B) and $\boldsymbol{\mu}_{AB}$ arrows, and for resonator B, it includes $\boldsymbol{\mu}_{AB}$, $1/j\omega$ (from resonator A) and $\boldsymbol{\mu}_{BA}$ arrows. As the SFG in Figure 2(c) illustrates, the mode coupling between the resonators also creates extra input-coupling routes for each of them, which leads to further modification of the input-coupling matrix for them. For resonator A, this input-coupling route includes $\boldsymbol{\kappa}^{B(L)}$, $1/j\omega$ (from resonator B) and $\boldsymbol{\mu}_{AB}$ arrows, and for resonator B, it includes $\boldsymbol{\kappa}^{A(L)}$, $1/j\omega$ (from resonator A) and $\boldsymbol{\mu}_{BA}$ arrows. Thus, direct coupling between the modes of the two resonators is, in fact, equivalent to the resonant loading effect of the principal modes that further modifies the pole and input-coupling matrices for each resonator. The resonant loading effect becomes more apparent when we decouple the CM equations, Eqs. (44) and (48), to get:

$$j\omega|\alpha\rangle=\left[\mathbf{P}_A^{(bl)}+\boldsymbol{\mu}_{AB}\left(j\omega\mathbf{I}-\mathbf{P}_B^{(bl)}\right)^{-1}\boldsymbol{\mu}_{BA}\right]|\alpha\rangle+\left[\boldsymbol{\kappa}^{A(bl)}+\boldsymbol{\mu}_{AB}\left(j\omega\mathbf{I}-\mathbf{P}_B^{(bl)}\right)^{-1}\boldsymbol{\kappa}^{B(bl)}\right]|S_{inc}\rangle, \tag{50a}$$

$$j\omega|\beta\rangle=\left[\mathbf{P}_B^{(bl)}+\boldsymbol{\mu}_{BA}\left(j\omega\mathbf{I}-\mathbf{P}_A^{(bl)}\right)^{-1}\boldsymbol{\mu}_{AB}\right]|\beta\rangle+\left[\boldsymbol{\kappa}^{B(bl)}+\boldsymbol{\mu}_{BA}\left(j\omega\mathbf{I}-\mathbf{P}_A^{(bl)}\right)^{-1}\boldsymbol{\kappa}^{A(bl)}\right]|S_{inc}\rangle. \tag{50b}$$

An alternative way to decouple the CM equations is to apply the SFG decomposition rules [4, 32] to Figure 2(c). From Eqs. (50), it becomes apparent that the non-resonant and resonant loading effects are responsible for the resonance frequency shifting that occurs upon the resonators' coupling. We define $A^{dl}$, the decoupled-loaded resonator A, as the resonator with the pole matrix $\mathbf{P}_A^{(dl)}=\mathbf{P}_A^{(bl)}+\boldsymbol{\mu}_{AB}\left(j\omega\mathbf{I}-\mathbf{P}_B^{(bl)}\right)^{-1}\boldsymbol{\mu}_{BA}$ and the input-coupling matrix $\boldsymbol{\kappa}^{A(dl)}=\boldsymbol{\kappa}^{A(bl)}+\boldsymbol{\mu}_{AB}\left(j\omega\mathbf{I}-\mathbf{P}_B^{(bl)}\right)^{-1}\boldsymbol{\kappa}^{B(bl)}$. Eq. (50a) is, in fact, the CM equation for the resonator $A^{dl}$ which can be considered as a black-box that captures the total loading effect of resonator B on resonator A. We define the resonator $B^{dl}$ same as $A^{dl}$. In terms of mode amplitudes and resonance frequencies, A and B coupled-resonator system is equivalent to system of the decoupled loaded resonators, $A^{dl}$ and $B^{dl}$ (Figure 2(e)). Hence, one can find the mode amplitudes by applying Eq. (31) to the resonators $A^{dl}$ and $B^{dl}$. The resonance frequencies can also be found by solving the characteristic equations for $A^{dl}$ and $B^{dl}$.

### 3.2. Energy, scattering, and absorption for the coupled resonators

We are now going to find the output power wave vector for the collective system of the resonators. By following the procedure used to derive Eq. (19) together with Eqs. (37) and (41), we get:

$$\begin{bmatrix} |S_{A1}^{-A}\rangle \\ |S_{B2}^{-B}\rangle \end{bmatrix} = \mathbf{S}^b \begin{bmatrix} |S_{A1}^{+A}\rangle \\ |S_{B2}^{+B}\rangle \end{bmatrix} + \begin{bmatrix} \mathbf{F}_{A1}^A \\ \mathbf{F}_{B2}^A \end{bmatrix} |\alpha\rangle + \begin{bmatrix} \mathbf{F}_{A1}^B \\ \mathbf{F}_{B2}^B \end{bmatrix} |\beta\rangle = \mathbf{S}^b |S_{inc}\rangle + \mathbf{F} \begin{bmatrix} |\alpha\rangle \\ |\beta\rangle \end{bmatrix}, \tag{51}$$

which has a similar form as Eq. (19). In Eq. (51), $\mathbf{S}^b$ is the scattering matrix for the collective system of the background structures, the elements of which can be acquired from numerical simulations, experimental measurements, or analytically by solving the integral equations (42). Also, $\mathbf{F} = \begin{bmatrix} \mathbf{F}^A & \mathbf{F}^B \end{bmatrix} = \begin{bmatrix} \mathbf{F}_{A1}^A & \mathbf{F}_{A1}^B \\ \mathbf{F}_{B2}^A & \mathbf{F}_{B2}^B \end{bmatrix}$ is the collective output-coupling matrix of the system modes with the unitless elements:

$$\mathbf{F}^A = \begin{bmatrix} \mathbf{F}_{A1}^A \\ \mathbf{F}_{B2}^A \end{bmatrix} = \frac{\mathbf{S}^f}{U_{nn}} \left[ v_{nm} = \left(-\frac{j\omega}{4}\right) \iiint_{space} \left[ \Delta\varepsilon_A \cdot \left(\tilde{E}_{Am} + \tilde{E}_{Am}^{sca(o)}\right) + \Delta\varepsilon_B \tilde{E}_{Am}^{sca(e)} \right] \cdot \left(E_n^w\right)^* e^{j\beta_n(z-z_{q_n})} d^3r \right]_{N \times M}$$

$$= \begin{bmatrix} \mathbf{F}_{A1}^{A(0)} \\ \mathbf{F}_{B2}^{A(0)} \end{bmatrix} + \frac{\mathbf{S}^f}{U_{nn}} \left[ v_{nm} = \left(-\frac{j\omega}{4}\right) \iiint_{\varepsilon_A \neq \varepsilon_{med}} \Delta\varepsilon_A \cdot \tilde{E}_{Am}^{sca(o)} \cdot \left(E_n^w\right)^* e^{j\beta_n(z-z_{q_n})} d^3r \right]_{N \times M} \tag{52a}$$

$$+ \frac{\mathbf{S}^f}{U_{nn}} \left[ v_{nm} = \left(-\frac{j\omega}{4}\right) \iiint_{\varepsilon_B \neq \varepsilon_{med}} \Delta\varepsilon_B \cdot \tilde{E}_{Am}^{sca(e)} \cdot \left(E_n^w\right)^* e^{j\beta_n(z-z_{q_n})} d^3r \right]_{N \times M},$$

$$\mathbf{F}^B = \begin{bmatrix} \mathbf{F}_{A1}^B \\ \mathbf{F}_{B2}^B \end{bmatrix} = \frac{\mathbf{S}^f}{U_{nn}} \left[ v_{nk} = \left(-\frac{j\omega}{4}\right) \iiint_{space} \left[ \Delta\varepsilon_B \cdot \left(\tilde{E}_{Bk} + \tilde{E}_{Bk}^{sca(o)}\right) + \Delta\varepsilon_A \tilde{E}_{Bk}^{sca(e)} \right] \cdot \left(E_n^w\right)^* e^{j\beta_n(z-z_{q_n})} d^3r \right]_{N \times K}$$

$$= \begin{bmatrix} \mathbf{F}_{A1}^{B(0)} \\ \mathbf{F}_{B2}^{B(0)} \end{bmatrix} + \frac{\mathbf{S}^f}{U_{nn}} \left[ v_{nk} = \left(-\frac{j\omega}{4}\right) \iiint_{\varepsilon_B \neq \varepsilon_{med}} \Delta\varepsilon_B \cdot \tilde{E}_{Bk}^{sca(o)} \cdot \left(E_n^w\right)^* e^{j\beta_n(z-z_{q_n})} d^3r \right]_{N \times K} \tag{52b}$$

$$+ \frac{\mathbf{S}^f}{U_{nn}} \left[ v_{nk} = \left(-\frac{j\omega}{4}\right) \iiint_{\varepsilon_A \neq \varepsilon_{med}} \Delta\varepsilon_A \cdot \tilde{E}_{Bk}^{sca(e)} \cdot \left(E_n^w\right)^* e^{j\beta_n(z-z_{q_n})} d^3r \right]_{N \times K},$$

where $\mathbf{S}^f$ is the scattering matrix of the free space slab between A1 and B2 reference planes and $q_n$ definition is similar to that in Eq. (45).

To find the collective stored energy and absorption for the coupled resonator system, we assume to be in the weak-coupling regime, i.e., when the high order scattered fields inside each resonator are much smaller than the normalized field patterns of the modes. Then, by following the procedures used to derive Eqs. (8), (15), and (21), we can obtain:

$$W \approx \text{Re}\left[\langle\alpha|\mathbf{H}_\mathbf{A}|\alpha\rangle\right] + \text{Re}\left[\langle\beta|\mathbf{H}_\mathbf{B}|\beta\rangle\right], \tag{53}$$

$$A \approx A_b - 2\omega\text{Im}\left[\langle\alpha|\mathbf{H}_\mathbf{A}|\alpha\rangle\right] - 2\omega\text{Im}\left[\langle\beta|\mathbf{H}_\mathbf{B}|\beta\rangle\right] = \\ \langle S_{inc}|S_{inc}\rangle - \langle S_{inc}|\left(\mathbf{S}^b\right)^\dagger \mathbf{S}^b|S_{inc}\rangle - 2\omega\text{Im}\left[\langle\alpha|\mathbf{H}_\mathbf{A}|\alpha\rangle\right] - 2\omega\text{Im}\left[\langle\beta|\mathbf{H}_\mathbf{B}|\beta\rangle\right], \tag{54}$$

with $A_b$ being the absorption in the collective system of background structures. When the resonators' modes are orthogonal and spectrally narrow, Eqs. (53) and (54) reduce to $W = \langle\alpha|\alpha\rangle + \langle\beta|\beta\rangle$ and

$A \approx A_b - \langle \alpha | 2\mathbf{\Gamma}_\mathbf{A}^{nr} | \alpha \rangle - \langle \beta | 2\mathbf{\Gamma}_\mathbf{B}^{nr} | \beta \rangle$, respectively. From Eqs. (53) and (54), it is deduced that the Hamiltonian matrices of the resonators are preserved in the weak-coupling regime, which for the simplified form of Eq. (54), further reduces into the preservation of the nonradiative decay rates for the modes inside the resonators. On the other hand, from Eqs. (50), we can see that coupling between the resonators changes the complex resonance frequencies of their modes. From these two results, one can deduce that in the weak-coupling regime, the resonators' mutual loading effect only modulates the real part of resonance frequencies and the radiative decay rates of their modes, which can be interpreted as the Purcell effect [5] on electromagnetic resonators.

The different coupling matrices are not entirely independent of each other. From Eqs. (44), (48), (51), and (54), and by following the same procedure employed for the derivation of Eq. (22), we can deduce that [35]:

$$\mathbf{F}^\dagger \mathbf{F} \approx 2\mathbf{\Gamma}^r - (\boldsymbol{\mu} + \boldsymbol{\mu}^\dagger), \tag{55}$$

with $\mathbf{\Gamma}^r = \begin{bmatrix} \mathbf{\Gamma}_\mathbf{A}^r & \mathbf{0} \\ \mathbf{0} & \mathbf{\Gamma}_\mathbf{B}^r \end{bmatrix}$. In the limit of the weak-coupling regime, Eq. (55) can be broken into the following two simpler equations [35]:

$$\boldsymbol{\mu}^r + (\boldsymbol{\mu}^r)^\dagger \approx 2\mathbf{\Gamma}^r - \mathbf{F}^\dagger \mathbf{F}, \tag{56a}$$

$$\boldsymbol{\mu}^n + (\boldsymbol{\mu}^n)^\dagger = 0, \tag{56b}$$

where $\boldsymbol{\mu}^r = \begin{bmatrix} \boldsymbol{\mu}_{AA}^r & \boldsymbol{\mu}_{AB}^r \\ \boldsymbol{\mu}_{BA}^r & \boldsymbol{\mu}_{BB}^r \end{bmatrix}$ and $\boldsymbol{\mu}^n = \begin{bmatrix} \boldsymbol{\mu}_{AA}^n & \boldsymbol{\mu}_{AB}^n \\ \boldsymbol{\mu}_{BA}^n & \boldsymbol{\mu}_{BB}^n \end{bmatrix}$. Eqs. (56) represent a generalization to the coupling coefficients relationship in TCMT formulation of coupled resonators [22].

### 3.3. Non-resonant background approximation

Suppose that the background structures of the resonators are off-resonant in the operating frequency range of interest. This condition causes the background structures to have a weak capability to store electromagnetic energy and, thus, allows us to approximate that the background structures scatter the propagating incident fields as propagating fields, and evanescent incident fields as evanescent fields. If the resonators are further separated in the propagation direction of waveguide modes (z-direction in Figure 2(a)), we can estimate their coupling terms from the uncoupled resonators' parameters. Based on the system geometry and from Eqs. (19), (41) and (42), we can write the following equations for the scattering of fields between the two resonators:

$$\begin{bmatrix} |S_{A1}^{-A}\rangle \\ |S_{A2}^{-A}\rangle \end{bmatrix} = \mathbf{S}^{Ab} \begin{bmatrix} |S_{A1}^{+A}\rangle \\ |S_{A2}^{+A}\rangle \end{bmatrix} + \begin{bmatrix} \mathbf{F}_{A1}^{A(0)} \\ \mathbf{F}_{A2}^{A(0)} \end{bmatrix} |\alpha\rangle = \begin{bmatrix} \mathbf{S}_{11}^{Ab} & \mathbf{S}_{12}^{Ab} \\ \mathbf{S}_{21}^{Ab} & \mathbf{S}_{22}^{Ab} \end{bmatrix} \begin{bmatrix} |S_{A1}^{+A}\rangle \\ |S_{A2}^{+A}\rangle \end{bmatrix} + \begin{bmatrix} \mathbf{F}_{A1}^{A(0)} \\ \mathbf{F}_{A2}^{A(0)} \end{bmatrix} |\alpha\rangle, \tag{57a}$$

$$\begin{bmatrix} \left|S_{B1}^{-B}\right\rangle \\ \left|S_{B2}^{-B}\right\rangle \end{bmatrix} = \mathbf{S}^{Bb} \begin{bmatrix} \left|S_{B1}^{+B}\right\rangle \\ \left|S_{B2}^{+B}\right\rangle \end{bmatrix} + \begin{bmatrix} \mathbf{F}_{B1}^{B(0)} \\ \mathbf{F}_{B2}^{B(0)} \end{bmatrix} |\beta\rangle = \begin{bmatrix} \mathbf{S}_{11}^{Bb} & \mathbf{S}_{12}^{Bb} \\ \mathbf{S}_{21}^{Bb} & \mathbf{S}_{22}^{Bb} \end{bmatrix} \begin{bmatrix} \left|S_{B1}^{+B}\right\rangle \\ \left|S_{B2}^{+B}\right\rangle \end{bmatrix} + \begin{bmatrix} \mathbf{F}_{B1}^{B(0)} \\ \mathbf{F}_{B2}^{B(0)} \end{bmatrix} |\beta\rangle, \quad (57b)$$

$$\left|S_{A2}^{+A}\right\rangle = \mathbf{D}(z_{B1} - z_{A2})\left|S_{B1}^{-B}\right\rangle, \quad (57c)$$

$$\left|S_{B1}^{+B}\right\rangle = \mathbf{D}(z_{B1} - z_{A2})\left|S_{A2}^{-A}\right\rangle, \quad (57d)$$

where $\left|S_{A2}^{+A}\right\rangle, \left|S_{B1}^{+B}\right\rangle, \left|S_{A1}^{-A}\right\rangle$ and $\left|S_{B2}^{+B}\right\rangle$ are the unknown terms. Eqs. (57a) and (57b) represent the formulas for the output power wave vectors of the two resonators, same as Eq. (19). On the other hand, Eqs. (57) and (57) describe the multiple scattering of the propagating fields between the two resonators.

By solving Eqs. (57) for $\left|S_{A2}^{+A}\right\rangle$ and $\left|S_{B1}^{+B}\right\rangle$, we can obtain:

$$\left|S_{A2}^{+A}\right\rangle = \mathbf{D}(z_{B1} - z_{A2})\left[\mathbf{I} - \mathbf{S}_{11}^{Bb}\mathbf{D}(z_{B1} - z_{A2})\mathbf{S}_{22}^{Ab}\mathbf{D}(z_{B1} - z_{A2})\right]^{-1}\left[\mathbf{S}_{11}^{Bb}\mathbf{D}(z_{B1} - z_{A2})\left|S_{A2}^{-A(0)}\right\rangle + \left|S_{B1}^{-B(0)}\right\rangle\right], \quad (58a)$$

$$\left|S_{B1}^{+B}\right\rangle = \mathbf{D}(z_{B1} - z_{A2})\left[\mathbf{I} - \mathbf{S}_{22}^{Ab}\mathbf{D}(z_{B1} - z_{A2})\mathbf{S}_{11}^{Bb}\mathbf{D}(z_{B1} - z_{A2})\right]^{-1}\left[\left|S_{A2}^{-A(0)}\right\rangle + \mathbf{S}_{22}^{Ab}\mathbf{D}(z_{B1} - z_{A2})\left|S_{B1}^{-B(0)}\right\rangle\right], \quad (58b)$$

where $\left|S_{A2}^{-A(0)}\right\rangle = \mathbf{S}_{21}^{Ab}\left|S_{A1}^{+A}\right\rangle + \mathbf{F}_{A2}^{A(0)}|\alpha\rangle$ is the 0$^{th}$ order output power wave vector of resonator A in region A2 when the incident field is from the region A1. Also, $\left|S_{B1}^{-B(0)}\right\rangle = \mathbf{S}_{12}^{Bb}\left|S_{B2}^{+B}\right\rangle + \mathbf{F}_{B1}^{B(0)}|\beta\rangle$ is the 0$^{th}$ order output power wave vector of resonator B in region B1 when the incident field is from the region B2.

By decomposing $E_B^{sca} = E_B^{sca,p} + E_B^{sca,np}$ in the right-hand side of Eq. (43) together with Eqs. (18), (57) and (58), we can estimate the radiative components of the input-coupling, self-coupling, and mode-coupling matrices (Eqs. (45)-(47)) in terms of the uncoupled resonators' parameters as:

$$\boldsymbol{\mu}_{AA}^r = \boldsymbol{\kappa}_{A2}^{A(0)}\mathbf{D}(z_{B1} - z_{A2})\left[\mathbf{I} - \mathbf{S}_{11}^{Bb}\mathbf{D}(z_{B1} - z_{A2})\mathbf{S}_{22}^{Ab}\mathbf{D}(z_{B1} - z_{A2})\right]^{-1}\mathbf{S}_{11}^{Bb}\mathbf{D}(z_{B1} - z_{A2})\mathbf{F}_{A2}^{A(0)}, \quad (59a)$$

$$\boldsymbol{\kappa}_{A1}^{A(bl)} = \boldsymbol{\kappa}_{A1}^{A(0)} + \boldsymbol{\kappa}_{A2}^{A(0)}\mathbf{D}(z_{B1} - z_{A2})\left[\mathbf{I} - \mathbf{S}_{11}^{Bb}\mathbf{D}(z_{B1} - z_{A2})\mathbf{S}_{22}^{Ab}\mathbf{D}(z_{B1} - z_{A2})\right]^{-1}\mathbf{S}_{11}^{Bb}\mathbf{D}(z_{B1} - z_{A2})\mathbf{S}_{21}^{Ab}, \quad (59b)$$

$$\boldsymbol{\kappa}_{B2}^{A(bl)} = \boldsymbol{\kappa}_{A2}^{A(0)}\mathbf{D}(z_{B1} - z_{A2})\left[\mathbf{I} - \mathbf{S}_{11}^{Bb}\mathbf{D}(z_{B1} - z_{A2})\mathbf{S}_{22}^{Ab}\mathbf{D}(z_{B1} - z_{A2})\right]^{-1}\mathbf{S}_{12}^{Bb}, \quad (59c)$$

$$\boldsymbol{\mu}_{AB}^r = \boldsymbol{\kappa}_{A2}^{A(0)}\mathbf{D}(z_{B1} - z_{A2})\left[\mathbf{I} - \mathbf{S}_{11}^{Bb}\mathbf{D}(z_{B1} - z_{A2})\mathbf{S}_{22}^{Ab}\mathbf{D}(z_{B1} - z_{A2})\right]^{-1}\mathbf{F}_{B1}^{B(0)}, \quad (59d)$$

with $\boldsymbol{\kappa}_{A1}^{A(0)}$ and $\boldsymbol{\kappa}_{A2}^{A(0)}$ being similar to Eq. (17). Similar equations as Eqs. (59) exist for $\boldsymbol{\mu}_{BB}^r$, $\boldsymbol{\kappa}_{A1}^{B(bl)}$, $\boldsymbol{\kappa}_{B2}^{B(bl)}$, and $\boldsymbol{\mu}_{BA}^r$. From Eqs. (59), we can rewrite the CM equations (44) and (48) as:

$$j\omega|\alpha\rangle = \left(\mathbf{P}_\mathbf{A}^{(0)} + \boldsymbol{\mu}_\mathbf{AA}^\mathbf{n} + \boldsymbol{\mu}_\mathbf{AA}^\mathbf{r}\right)|\alpha\rangle + \boldsymbol{\kappa}_\mathbf{A1}^{\mathbf{A(bl)}}|S_{A1}^{+A}\rangle + \boldsymbol{\kappa}_\mathbf{B2}^{\mathbf{A(bl)}}|S_{B2}^{+B}\rangle + \left(\boldsymbol{\mu}_\mathbf{AB}^\mathbf{n} + \boldsymbol{\mu}_\mathbf{AB}^\mathbf{r}\right)|\beta\rangle$$
$$= \left(\mathbf{P}_\mathbf{A}^{(0)} + \boldsymbol{\mu}_\mathbf{AA}^\mathbf{n}\right)|\alpha\rangle + \boldsymbol{\kappa}_\mathbf{A1}^{\mathbf{A(0)}}|S_{A1}^{+A}\rangle + \boldsymbol{\kappa}_\mathbf{A2}^{\mathbf{A(0)}}|S_{A2}^{+A}\rangle + \boldsymbol{\mu}_\mathbf{AB}^\mathbf{n}|\beta\rangle, \tag{60a}$$

$$j\omega|\beta\rangle = \left(\mathbf{P}_\mathbf{B}^{(0)} + \boldsymbol{\mu}_\mathbf{BB}^\mathbf{n} + \boldsymbol{\mu}_\mathbf{BB}^\mathbf{r}\right)|\beta\rangle + \boldsymbol{\kappa}_\mathbf{A1}^{\mathbf{B(bl)}}|S_{A1}^{+A}\rangle + \boldsymbol{\kappa}_\mathbf{B2}^{\mathbf{B(bl)}}|S_{B2}^{+B}\rangle + \left(\boldsymbol{\mu}_\mathbf{BA}^\mathbf{n} + \boldsymbol{\mu}_\mathbf{BA}^\mathbf{r}\right)|\alpha\rangle$$
$$= \left(\mathbf{P}_\mathbf{B}^{(0)} + \boldsymbol{\mu}_\mathbf{BB}^\mathbf{n}\right)|\beta\rangle + \boldsymbol{\kappa}_\mathbf{B1}^{\mathbf{B(0)}}|S_{B1}^{+B}\rangle + \boldsymbol{\kappa}_\mathbf{B2}^{\mathbf{B(0)}}|S_{B2}^{+B}\rangle + \boldsymbol{\mu}_\mathbf{BA}^\mathbf{n}|\alpha\rangle, \tag{60b}$$

which are similar to the CM equations for the uncoupled resonators only with the extra near-field coupling terms $\boldsymbol{\mu}_\mathbf{AB}^\mathbf{n}$ and $\boldsymbol{\mu}_\mathbf{BA}^\mathbf{n}$.

By solving Eqs. (57) for $|S_{A1}^{-A}\rangle$ and $|S_{B2}^{-B}\rangle$, we can also estimate the scattering matrix elements for the collective system of the background structures, as well as the output-coupling matrix elements (in Eq. (51)) in terms of the uncoupled resonators' parameters:

$$\mathbf{S}_\mathbf{11}^\mathbf{b} = \mathbf{S}_\mathbf{11}^\mathbf{Ab} + \mathbf{S}_\mathbf{12}^\mathbf{Ab}\mathbf{D}(z_{B1}-z_{A2})\left[\mathbf{I}-\mathbf{S}_\mathbf{11}^\mathbf{Bb}\mathbf{D}(z_{B1}-z_{A2})\mathbf{S}_\mathbf{22}^\mathbf{Ab}\mathbf{D}(z_{B1}-z_{A2})\right]^{-1}\mathbf{S}_\mathbf{11}^\mathbf{Bb}\mathbf{D}(z_{B1}-z_{A2})\mathbf{S}_\mathbf{21}^\mathbf{Ab}, \tag{61a}$$

$$\mathbf{S}_\mathbf{12}^\mathbf{b} = \mathbf{S}_\mathbf{12}^\mathbf{Ab}\mathbf{D}(z_{B1}-z_{A2})\left[\mathbf{I}-\mathbf{S}_\mathbf{11}^\mathbf{Bb}\mathbf{D}(z_{B1}-z_{A2})\mathbf{S}_\mathbf{22}^\mathbf{Ab}\mathbf{D}(z_{B1}-z_{A2})\right]^{-1}\mathbf{S}_\mathbf{12}^\mathbf{Bb}, \tag{61b}$$

$$\mathbf{S}_\mathbf{21}^\mathbf{b} = \mathbf{S}_\mathbf{21}^\mathbf{Bb}\mathbf{D}(z_{B1}-z_{A2})\left[\mathbf{I}-\mathbf{S}_\mathbf{22}^\mathbf{Ab}\mathbf{D}(z_{B1}-z_{A2})\mathbf{S}_\mathbf{11}^\mathbf{Bb}\mathbf{D}(z_{B1}-z_{A2})\right]^{-1}\mathbf{S}_\mathbf{21}^\mathbf{Ab}, \tag{61c}$$

$$\mathbf{S}_\mathbf{22}^\mathbf{b} = \mathbf{S}_\mathbf{22}^\mathbf{Bb} + \mathbf{S}_\mathbf{21}^\mathbf{Bb}\mathbf{D}(z_{B1}-z_{A2})\left[\mathbf{I}-\mathbf{S}_\mathbf{22}^\mathbf{Ab}\mathbf{D}(z_{B1}-z_{A2})\mathbf{S}_\mathbf{11}^\mathbf{Bb}\mathbf{D}(z_{B1}-z_{A2})\right]^{-1}\mathbf{S}_\mathbf{22}^\mathbf{Ab}\mathbf{D}(z_{B1}-z_{A2})\mathbf{S}_\mathbf{12}^\mathbf{Bb}, \tag{61d}$$

$$\mathbf{F}_\mathbf{A1}^\mathbf{A} = \mathbf{F}_\mathbf{A1}^\mathbf{A(0)} + \mathbf{S}_\mathbf{12}^\mathbf{Ab}\mathbf{D}(z_{B1}-z_{A2})\left[\mathbf{I}-\mathbf{S}_\mathbf{11}^\mathbf{Bb}\mathbf{D}(z_{B1}-z_{A2})\mathbf{S}_\mathbf{22}^\mathbf{Ab}\mathbf{D}(z_{B1}-z_{A2})\right]^{-1}\mathbf{S}_\mathbf{11}^\mathbf{Bb}\mathbf{D}(z_{B1}-z_{A2})\mathbf{F}_\mathbf{A2}^\mathbf{A(0)}, \tag{61e}$$

$$\mathbf{F}_\mathbf{B2}^\mathbf{A} = \mathbf{S}_\mathbf{21}^\mathbf{Bb}\mathbf{D}(z_{B1}-z_{A2})\left[\mathbf{I}-\mathbf{S}_\mathbf{22}^\mathbf{Ab}\mathbf{D}(z_{B1}-z_{A2})\mathbf{S}_\mathbf{11}^\mathbf{Bb}\mathbf{D}(z_{B1}-z_{A2})\right]^{-1}\mathbf{F}_\mathbf{A2}^\mathbf{A(0)}, \tag{61f}$$

$$\mathbf{F}_\mathbf{A1}^\mathbf{B} = \mathbf{S}_\mathbf{12}^\mathbf{Ab}\mathbf{D}(z_{B1}-z_{A2})\left[\mathbf{I}-\mathbf{S}_\mathbf{11}^\mathbf{Bb}\mathbf{D}(z_{B1}-z_{A2})\mathbf{S}_\mathbf{22}^\mathbf{Ab}\mathbf{D}(z_{B1}-z_{A2})\right]^{-1}\mathbf{F}_\mathbf{B1}^\mathbf{B(0)}, \tag{61g}$$

$$\mathbf{F}_\mathbf{B2}^\mathbf{B} = \mathbf{F}_\mathbf{B2}^\mathbf{B(0)} + \mathbf{S}_\mathbf{21}^\mathbf{Bb}\mathbf{D}(z_{B1}-z_{A2})\left[\mathbf{I}-\mathbf{S}_\mathbf{22}^\mathbf{Ab}\mathbf{D}(z_{B1}-z_{A2})\mathbf{S}_\mathbf{11}^\mathbf{Bb}\mathbf{D}(z_{B1}-z_{A2})\right]^{-1}\mathbf{S}_\mathbf{22}^\mathbf{Ab}\mathbf{D}(z_{B1}-z_{A2})\mathbf{F}_\mathbf{B1}^\mathbf{B(0)}. \tag{61h}$$

Eqs. (58) and (61) can also be obtained from Eqs. (42).

Finally, in the regime of non-resonant background approximation, we can estimate $A_b$, i.e., the absorption term for the collective system of the background structures in Eq. (54) as:

$$A_b \approx \langle S^{+A}|S^{+A}\rangle - \langle S^{+A}|\left(\mathbf{S}^\mathbf{Ab}\right)^\dagger \mathbf{S}^\mathbf{Ab}|S^{+A}\rangle + \langle S^{+B}|S^{+B}\rangle - \langle S^{+B}|\left(\mathbf{S}^\mathbf{Bb}\right)^\dagger \mathbf{S}^\mathbf{Bb}|S^{+B}\rangle, \tag{62}$$

where $\left|S^{+A}\right\rangle = \begin{bmatrix} \left|S_{A1}^{+A}\right\rangle \\ \left|S_{A2}^{+A}\right\rangle \end{bmatrix}$ and $\left|S^{+B}\right\rangle = \begin{bmatrix} \left|S_{B1}^{+B}\right\rangle \\ \left|S_{B2}^{+B}\right\rangle \end{bmatrix}$.

In summary, we thoroughly analyzed the phenomenon of coupling between resonators using the BBCMT framework in section 3. Although we only focused on the coupling between two resonators for simplicity, the approach is similar when coupling among three or more resonators is considered. BBCMT framework for coupled resonators reduces to that for TCMT when single-mode, closed, dispersionless, and lossless resonators are assumed. As the final remark on this section, we mention the particular case when all of the coupled resonators, except one of them (e.g. resonator A), are non-resonant in the operating frequency range. Under such condition, BBCMT equations reduce to the description of resonator A's behavior when loaded by other non-resonant scatterers. In this regard, BBCMT could also be considered as a generalization of the *cavity perturbation theory* [19, 30, 34], which analyzes the shifting in resonance frequencies of a resonator perturbed by non-resonant scatterers in its neighboring.

## 4. Applications and numerical validation

To signify our theory's validity, we consider two examples in which the far-field optical spectra obtained from BBCMT models are compared to those obtained from finite difference time domain (FDTD) simulations. The example in section 4.1 demonstrates the application of the background perturbation method to express the scattering and absorption optical responses for a single resonator. On the other hand, the example in section 4.2 shows the importance of considering the non-resonant loading effect to accurately model the optical responses for a coupled resonator system. In addition to BBCMT's numerical validation, both examples also exhibit the BBCMT generality by closely comparing its predictions to the conventional TCMT models' predictions.

### 4.1. Appropriate background structure for a resonator

Figure 3(a) shows a metal-insulator-metal (MIM) silver (Ag) nanodisk array embedded in an $SiO_2$ slab (configuration details in SI section 4.1 [35]), which is illuminated by an x-polarized plane wave in a normal direction from the top. Reflectance, transmittance, and absorbance spectra of the system obtained from FDTD simulations are shown in Figures 3(b-d) (details in SI section 3 [35]). One can see that the system spectral response over the frequency range from $\omega = 1.8 \times 10^{15}$ rad/s to $\omega = 2.6 \times 10^{15}$ rad/s is dominated by a localized surface plasmon (LSP) mode with the resonance frequency of $\Omega_A = 2.19 \times 10^{15}$ rad/s and associated with the nanogap between the silver nanodisks (Figure S2 [35]). We then use BBCMT to model the spectral response of the system in the designated frequency range.

Rather than treating each MIM nanodisk as one resonator, we consider the MIM nanodisk array and its embedding $SiO_2$ layer to represent a single resonator A collectively. Since the array periodicity is smaller than the light wavelength in our spectral range of interest ($P = 400$ nm $< \lambda$), the phase-matching condition only allows for the $0^{th}$ order backscattering (or reflection) and forward-scattering (or transmission) of the normally-incident plane wave [3]. Moreover, the system does not convert x-polarization to y-polarization or vice versa. Therefore, we can take resonator A to only have two ports of top and bottom. We have considered the reference planes for each port to overlap with the embedding $SiO_2$ slab's surface. The resonator structural symmetry makes its modes orthogonal. Also, the resonator materials are low-dispersion (almost) and low-loss over the designated frequency range (SI section 3 [35]). According to these considerations and from Eqs. (9), (15), (17), (19), (22) and (23), we get the following equations for mode coupling, output power wave vector, and absorption of resonator A:

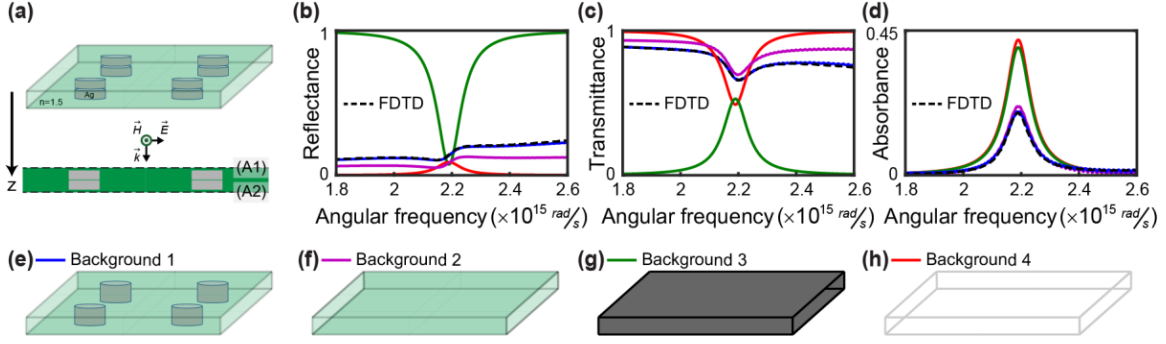

**Figure 3**. An example to clarify the concept of a proper background structure and numerically validate the background perturbation method. (a) The resonator geometry: An MIM nanodisk array embedded in an SiO$_2$ slab. (b) Reflectance, (c) transmittance, and (d) absorbance spectra over the designated frequency. The dashed lines represent the spectra calculated from the FDTD simulation. The solid lines represent the spectra calculated by BBCMT models with (e) background 1: a metal nanodisk array embedded in the SiO$_2$ slab, (f) background 2: the embedding SiO$_2$ slab, (g) background 3: a PEC ground plane slab, and (h) background 4: a free space slab. Backgrounds 3 and 4 correspond to the directly-coupled and the side-coupled TCMT versions, respectively.

$$j\omega\alpha = P_A^{(0)}\alpha + \begin{bmatrix} \kappa_{A1}^{A(0)} & \kappa_{A2}^{A(0)} \end{bmatrix} \begin{bmatrix} S_{A1}^{+A} \\ S_{A2}^{+A} \end{bmatrix}, \tag{63a}$$

$$\begin{bmatrix} S_{A1}^{-A} \\ S_{A2}^{-A} \end{bmatrix} = \mathbf{S}^{Ab} \begin{bmatrix} S_{A1}^{+A} \\ S_{A2}^{+A} \end{bmatrix} + \begin{bmatrix} f_{A1}^{A(0)} \\ f_{A2}^{A(0)} \end{bmatrix} \alpha, \tag{63b}$$

$$\begin{bmatrix} \kappa_{A1}^{A(0)} & \kappa_{A2}^{A(0)} \end{bmatrix} = -\begin{bmatrix} \left(f_{A1}^{A(0)}\right)^* & \left(f_{A2}^{A(0)}\right)^* \end{bmatrix} \mathbf{S}^{Ab}, \tag{63c}$$

$$A_A = 2\Gamma_A^{nr}|\alpha|^2, \tag{63d}$$

where $\alpha$ is the mode amplitude and $P_A^{(0)} = j\Omega_A - \Gamma_A^{nr} - \Gamma_A^r$ is the mode pole with $\Gamma_A^{nr}$ and $\Gamma_A^r$ being the nonradiative and radiative decay rates, respectively. $\begin{bmatrix} \kappa_{A1}^{A(0)} & \kappa_{A2}^{A(0)} \end{bmatrix}$ is the mode input-coupling matrix and $\begin{bmatrix} S_{A1}^{+A} \\ S_{A2}^{+A} \end{bmatrix}$ is the input power wave vector. Moreover, $\begin{bmatrix} f_{A1}^{A(0)} \\ f_{A2}^{A(0)} \end{bmatrix}$ is the mode output-coupling matrix and $\mathbf{S}^{Ab}$ is the scattering matrix of the resonator background structure. We did the BBCMT modeling with two background structures (Figure 3(e-f)): 1) A geometry similar to resonator A but with the gap filled with silver, i.e., a metal nanodisk array embedded in the SiO$_2$ slab; 2) The embedding SiO$_2$ slab of resonator A without the MIM nanodisks. For each background structure, we used the F-estimation lemma to acquire the resonator parameters in Eqs. (63) (SI section 4.1 [35]). Then, by putting $S_{A1}^{+A} = 1$ and $S_{A2}^{+A} = 0$, we calculated the reflectance, transmittance, and absorbance spectra of the system (Figures 3(b-d)). BBCMT-calculated spectra for background 2 correctly capture the overall shapes of the FDTD-calculated spectra but do not perfectly fit them. However, BBCMT-calculated spectra for background 1 match the FDTD-calculated results very well. Despite generating less accurate results than those from background 1, background 2 has the virtue of being less complex, leading to a reduced computational cost for finding the background scattering matrix. This example shows the trade-off between the accuracy and complexity in choosing the background structure for BBCMT models.

Two common TCMT versions are directly-coupled and side-coupled models. The directly-coupled version assumes a high non-resonant scattering from the resonator and thus can model the resonators based on dielectric slabs of high refractive index or metallic slabs [21, 22]. Directly-coupled TCMT model can also treat the direct-coupling processes between waveguides and resonators in photonic crystal structures [25]. On the other hand, the side-coupled version assumes a negligible non-resonant scattering from the resonator and thus is typically suitable to model the resonators much smaller than the operating wavelength [28]. The side-coupled TCMT model can also treat the side-coupling processes between waveguides and resonators in photonic crystal structures [23].

In addition to the background structures 1 and 2 that are used for modeling of the system in Figure 3(a) with BBCMT, we also perform the modeling for two other background structures which correspond to the directly-coupled and side-coupled TCMT models [35] as demonstrated in section 4.2: Background 3 is a perfect electric conductor (PEC) ground-plane slab (Figure 3(g)) that corresponds to the directly-coupled TCMT version [22]. Background 4 is a Free-space slab (Figure 3(h)) that corresponds to the side-coupled TCMT version [23]. Figures 3(b-d) reveal that both of these background structures are inappropriate for modeling the resonator in Figure 3(a). The reason is that the $SiO_2$ slab is neither as transparent as a free-space slab nor as reflective as a ground-plane but has a reflectivity between these two cases.

### 4.2. BBCMT-aided resonator design

BBCMT is a powerful tool to obtain a microscopic understanding of complex electromagnetic resonator systems, which can help their design. We demonstrate this capability by the following example. Figure 4(a) shows a nanolaminate MIM plasmonic crystal consisting of an array of tapered MIM nanodomes over a nanolaminate MIM nanohole array with tapered nanoholes (details in SI section 4.2 [35]). The incident wave is an x-polarized plane wave normally illuminated from the top. Our goal is to find the optimal distance between the nanodome and nanohole arrays ($d = z_{B1} - z_{A2}$ in Figure 4(a)), which maximizes the magnitude and bandwidth of light absorption in the frequency range of $\omega = 1.6 \times 10^{15}$ rad/s to $\omega = 2.2 \times 10^{15}$ rad/s. To achieve the goal, we first find the proper BBCMT models for the uncoupled nanodome and nanohole arrays and then build the BBCMT model for the plasmonic crystal by analyzing the coupling between the nanodome and nanohole arrays. Next, we extract the BBCMT model parameters by fitting the FDTD-simulated absorbance spectra of the uncoupled arrays and the plasmonic crystal for $d$=80 nm and $d$=120 nm. After that, we find the optimal value of $d$ using the microscopic picture acquired from the BBCMT analysis. Finally, we verify the design and analysis by comparing the reflectance, transmittance, and absorbance spectra that BBCMT predicts for $d$=0-2000 nm against the spectra obtained from FDTD simulations.

#### 4.2.1. BBCMT model of the system

We start by finding the BBCMT model for the uncoupled nanodome array (resonator A). As the nanodomes have subwavelength dimensions, their non-resonant absorption and scattering are negligible [28], and the incident light entirely transmits through the nanodome array unless it gets resonantly scattered or absorbed by the LSP modes in the system. One can understand this argument from the coincidence of absorbance peaks and transmittance dips for the nanodome array shown in Figure 4(b). Moreover, the array's periodicity is subwavelength ($P = 400\ nm < \lambda$), and it preserves the polarization of the incident wave over the frequency range of interest. Accordingly, we can take resonator A to have two-ports (top and bottom ports) and a free space background with the scattering matrix $\mathbf{S}^{Ab} = \mathbf{S}^{Af} = \begin{bmatrix} 0 & e^{-jkd_A} \\ e^{-jkd_A} & 0 \end{bmatrix}$, where $k = \omega/c = 2\pi/\lambda$ is the free-space wavenumber with $c$ being the speed of light in vacuum.

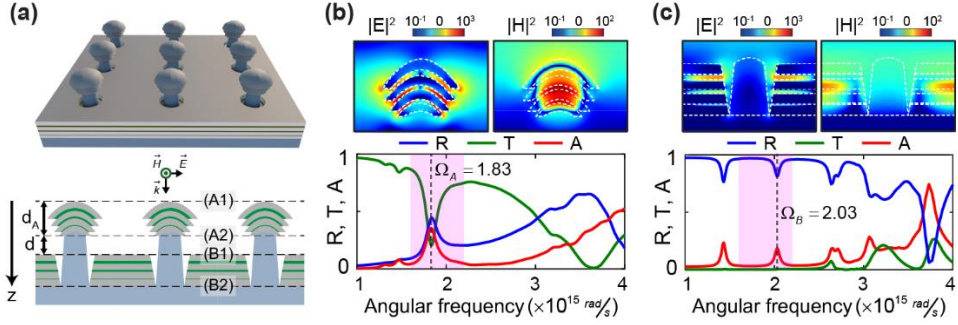

**Figure 4.** The nanolaminate plasmonic crystal and its constituent nanodome and nanohole arrays. (a) The geometry of the nanolaminate plasmonic crystal. $d$ represents the distance between the nanodome and nanohole arrays. The system is illuminated with an x-polarized plane wave incident from the top side. (b, c) The FDTD-simulated reflectance, transmittance, and absorbance spectra of the uncoupled nanodome array (b) and the uncoupled nanohole array (c). The figures also show the near-field distributions of the dominant mode for each subsystem (at resonance) in the designated frequency range of interest ($\omega = 1.6 \times 10^{15}$ rad/s to $\omega = 2.2 \times 10^{15}$ rad/s) shaded in purple.

Also, $d_A = |z_{A2} - z_{A1}| = 150$ nm displays the distance between the resonator reference planes in the top and bottom regions. The reference planes $z = z_{A1}$ and $z = z_{A2}$ are chosen to cross the top and bottom resonator points, respectively (Figure 4(a)).

Far-field spectra of resonator A reveals that the mode with the resonance frequency of $\Omega_A = 1.83 \times 10^{15}$ rad/s dominates the resonator response in the designated frequency range (purple-shaded area in Figure 4(b)). Near-field distributions of the components $|E|^2$ and $|H|^2$ for this mode (at $\omega = \Omega_A$) in Figure 4(b) indicate that it is a plasmonic gap mode with a dominant magnetic dipole nature [38, 40]. Just as the MIM nanodisk array in section 4.1, we consider the resonator modes to be orthogonal for simplicity. Then, from Eqs. (9), (15), (17), (19), (22) and (23), and considering the materials' small dispersion and loss, we get the following equations for resonator A:

$$j\omega\alpha = P_A^{(0)}\alpha + \begin{bmatrix} \kappa_{A1}^{A(0)} & \kappa_{A2}^{A(0)} \end{bmatrix} \begin{bmatrix} S_{A1}^{+A} \\ S_{A2}^{+A} \end{bmatrix} = \left(j\Omega_A - \Gamma_A^{nr} - \Gamma_A^{r,1} - \Gamma_A^{r,2}\right)\alpha + \kappa_{A1}^{A(0)} S_{A1}^{+A} + \kappa_{A2}^{A(0)} S_{A2}^{+A}, \quad (64a)$$

$$\begin{bmatrix} S_{A1}^{-A} \\ S_{A2}^{-A} \end{bmatrix} = \begin{bmatrix} 0 & e^{-jkd_A} \\ e^{-jkd_A} & 0 \end{bmatrix} \begin{bmatrix} S_{A1}^{+A} \\ S_{A2}^{+A} \end{bmatrix} - \begin{bmatrix} 0 & e^{-jkd_A} \\ e^{-jkd_A} & 0 \end{bmatrix} \begin{bmatrix} \left(\kappa_{A1}^{A(0)}\right)^* \\ \left(\kappa_{A2}^{A(0)}\right)^* \end{bmatrix} \alpha, \quad (64b)$$

$$\left|\kappa_{A1}^{A(0)}\right| = \sqrt{2\Gamma_A^{r,2}}, \left|\kappa_{A2}^{A(0)}\right| = \sqrt{2\Gamma_A^{r,1}}, \quad (64c)$$

$$A_A = 2\Gamma_A^{nr}|\alpha|^2, \quad (64d)$$

where $\alpha$ is the mode amplitude, $P_A^{(0)}$ is the pole matrix, and $\Gamma_A^{r,1}$ and $\Gamma_A^{r,2}$ represent the decay rates of the mode into the resonator top and bottom ports, respectively. We have considered the decay rates and the coupling coefficients to have the units of $1/s$ in Eq. (64c). Eqs. (64) match the equations derived for side-coupled resonators in other works [23].

Next, we find the BBCMT model for the MIM nanohole array. The nanohole array is essentially a planar MIM structure perforated with a regular array of subwavelength nanoholes. Such structures support several Bloch SPP modes in their gaps/interfaces [33]. The nanoholes' subwavelength size does not allow for the non-resonant transmission of the incident light through them. Therefore, when the incident light's frequency shifts away from the Bloch SPP modes' resonances, the nanohole array behaves almost exactly as the underlying planar MIM structure, and most of the incident light gets reflected. However, as shown in Figure 4(c), resonant excitation of the Bloch SPP modes causes dips in the reflectance spectra simultaneous with peaks in the absorbance spectra. We also get resonant peaks in the transmittance spectra when the Bloch SPP modes on the bottom interface get excited [36]. Moreover, the array's periodicity is subwavelength ($P = 400\,nm < \lambda$), and it preserves the polarization of the incident wave over the frequency range of interest. According to the above explanations, we can consider the nanohole array to be a two-port resonator B with the underlying planar MIM structure as the background. Since the planar MIM structure behaves almost like a PEC ground plane, we can write $\mathbf{S^{Bb}} \approx \mathbf{S}^{ground\ plane} = \begin{bmatrix} -1 & 0 \\ 0 & -1 \end{bmatrix}$, where the reference planes are chosen to overlap with the resonator's top and bottom surfaces (Figure 4(a)).

Far-field spectra of resonator B reveals that the mode with the resonance frequency of $\Omega_B = 2.03 \times 10^{15}$ rad/s dominates the resonator response in the designated frequency range (purple-shaded area in Figure 4(c)). Near-field distributions of this mode (at $\omega = \Omega_B$) in Figure 4(c) indicate that it is a Bloch SPP mode associated with the dielectric gaps. We consider the resonator modes to be orthogonal for simplicity. Then, from Eqs. (9), (15), (17), (19), (22) and (23), and in view of the materials' small dispersion and loss, we get the following equations for resonator B:

$$j\omega\beta = P_B^{(0)}\beta + \begin{bmatrix} \kappa_{B1}^{B(0)} & \kappa_{B2}^{B(0)} \end{bmatrix} \begin{bmatrix} S_{B1}^{+B} \\ S_{B2}^{+B} \end{bmatrix} = \left(j\Omega_B - \Gamma_B^{nr} - \Gamma_B^{r,1} - \Gamma_B^{r,2}\right)\beta + \kappa_{B1}^{B(0)} S_{B1}^{+B} + \kappa_{B2}^{B(0)} S_{B2}^{+B}, \quad (65a)$$

$$\begin{bmatrix} S_{B1}^{-B} \\ S_{B2}^{-B} \end{bmatrix} = \begin{bmatrix} -1 & 0 \\ 0 & -1 \end{bmatrix} \begin{bmatrix} S_{B1}^{+B} \\ S_{B2}^{+B} \end{bmatrix} - \begin{bmatrix} -1 & 0 \\ 0 & -1 \end{bmatrix} \begin{bmatrix} \left(\kappa_{B1}^{B(0)}\right)^* \\ \left(\kappa_{B2}^{B(0)}\right)^* \end{bmatrix} \beta, \quad (65b)$$

$$\left|\kappa_{B1}^{B(0)}\right| = \sqrt{2\Gamma_B^{r,1}}, \left|\kappa_{B2}^{B(0)}\right| = \sqrt{2\Gamma_B^{r,2}}, \quad (65c)$$

$$A_B = 2\Gamma_B^{nr}|\beta|^2, \quad (65d)$$

where $\beta$ is the mode amplitude, and $P_B^{(0)}$ is the pole matrix, and $\Gamma_B^{r,1}$ and $\Gamma_B^{r,2}$ are defined similar to those in Eqs. (64). Eqs. (65) match the equations derived for directly-coupled resonators in other works [22, 25].

The schematic in Figure 5(a) represents the BBCMT model for the nanolaminate plasmonic crystal, or equivalently, the system of resonators A and B coupled through near-field and radiative interactions. Here, we have neglected the effect of polymer pillars between the nanodome and nanohole arrays as they are both non-resonant and have a negligible scattering due to subwavelength cross-section. Since the background structures of the resonators A and B are non-resonant, and they do not overlap in the wave propagation direction, we may use the approximations in section 3.3. From Eq. (60), we get the following CM equations for the coupled system:

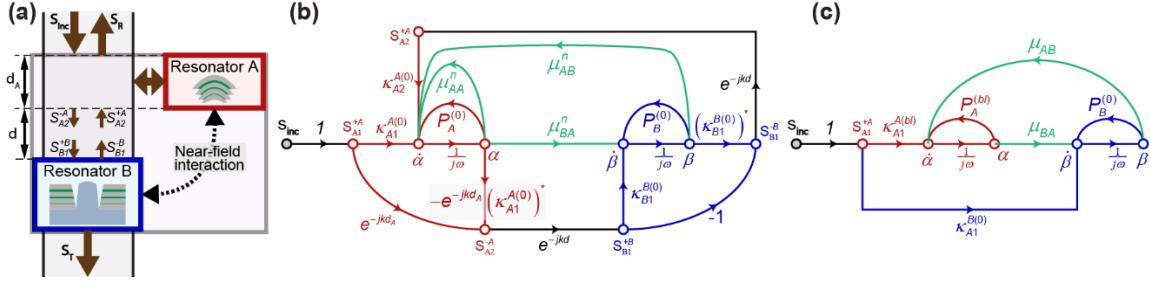

**Figure 5.** (a) BBCMT model for the nanolaminate plasmonic crystal. (b) The SFG of the system when we assume a free space background structure for the nanodome array and a background structure of the PEC ground plane for the nanohole array. (c) The SFG resulted from the simplification of the SFG in part (b) by the SFG decomposition rules.

$$j\omega\alpha = \left[P_A^{(0)} + \mu_{AA}^n\right]\alpha + \kappa_{A1}^{A(0)} S_{inc} + \kappa_{A2}^{A(0)} S_{A2}^{+A} + \mu_{AB}^n \beta, \tag{66a}$$

$$j\omega\beta = P_B^{(0)}\beta + \kappa_{B1}^{B(0)} S_{B1}^{+B} + \mu_{BA}^n \alpha, \tag{66b}$$

where we have considered $S_{B2}^{+B} = 0$ and $S_{A1}^{+A} = S_{inc}$ as the excitation wave is incident from the top side. Since resonator B's background structure is a ground plane, the self-coupling term $\mu_{AA}^n$ can be interpreted as the near-field interaction between resonator A and its image in the ground plane. As the non-resonant scattering from resonator A is negligible, we have not considered any self-coupling term for resonator B. Furthermore, considering that the magnitudes of near-field components for a resonator exponentially decay with the distance away from the resonator, we can approximate the near-field coupling terms in Eqs. (66) to be exponentially decaying functions of $d$:

$$\mu_{AA}^n = \mu_{AA}^{n0} e^{-\alpha_{AA}kd}, \tag{67a}$$

$$\mu_{AB}^n = \mu_{AB}^{n0} e^{-\alpha_{AB}kd}, \tag{67b}$$

$$\mu_{BA}^n = \mu_{BA}^{n0} e^{-\alpha_{BA}kd}, \tag{67c}$$

with $\mu_{AA}^{n0}$, $\mu_{AB}^{n0}$ and $\mu_{BA}^{n0}$ being the near-field coupling terms for $d=0$ and $\alpha_{AA}$, $\alpha_{AB}$ and $\alpha_{BA}$ being the normalized spatial decay rates.

From Eqs. (57), (64), and (65) and considering the geometry of the coupled system in Figure 5(a), we can write the following equations for the scattering of fields between the two resonators:

$$\begin{bmatrix} S_R \\ S_{A2}^{-A} \end{bmatrix} = \begin{bmatrix} 0 & e^{-jkd_A} \\ e^{-jkd_A} & 0 \end{bmatrix} \begin{bmatrix} S_{inc} \\ S_{A2}^{+A} \end{bmatrix} - \begin{bmatrix} 0 & e^{-jkd_A} \\ e^{-jkd_A} & 0 \end{bmatrix} \begin{bmatrix} \left(\kappa_{A1}^{A(0)}\right)^* \\ \left(\kappa_{A2}^{A(0)}\right)^* \end{bmatrix} \alpha, \tag{68a}$$

$$\begin{bmatrix} S_{B1}^{-B} \\ S_T \end{bmatrix} = \begin{bmatrix} -1 & 0 \\ 0 & -1 \end{bmatrix} \begin{bmatrix} S_{B1}^{+B} \\ 0 \end{bmatrix} - \begin{bmatrix} -1 & 0 \\ 0 & -1 \end{bmatrix} \begin{bmatrix} \left(\kappa_{B1}^{B(0)}\right)^* \\ \left(\kappa_{B2}^{B(0)}\right)^* \end{bmatrix} \beta, \tag{68b}$$

$$S_{A2}^{+A} = e^{-jkd} S_{B1}^{-B}, \tag{68c}$$

$$S_{B1}^{+B} = e^{-jkd} S_{A2}^{-A}, \tag{68d}$$

where $S_R = S_{A1}^{-A}$ and $S_T = S_{B2}^{-B}$ represent the reflected and transmitted power waves for the coupled system, respectively. The SFG in Figure 5(b) illustrates the electromagnetic interaction dynamics associated with Eqs. (66) and (68) c-d. Via applying the SFG decomposition rules [4, 32] to Figure 5(b), we can simplify it to the SFG in Figure 5(c), which corresponds to the following equations:

$$j\omega\alpha = P_A^{(bl)}\alpha + \kappa_{A1}^{A(bl)} S_{inc} + \mu_{AB}\beta \tag{69a}$$

$$j\omega\beta = P_B^{(0)}\beta + \kappa_{A1}^{B(0)} S_{inc} + \mu_{BA}\alpha, \tag{69b}$$

where $P_A^{(bl)} = P_A^{(0)} + \mu_{AA}^n + \mu_{AA}^r$, $\mu_{AB} = \mu_{AB}^n + \mu_{AB}^r$ and $\mu_{BA} = \mu_{BA}^n + \mu_{BA}^r$ with:

$$\mu_{AA}^r = \kappa_{A2}^{A(0)} e^{-jk(d_A+2d)} \left(\kappa_{A1}^{A(0)}\right)^*, \tag{70a}$$

$$\kappa_{A1}^{A(bl)} = \kappa_{A1}^{A(0)} - \kappa_{A2}^{A(0)} e^{-jk(d_A+2d)}, \tag{70b}$$

$$\mu_{AB}^r = \kappa_{A2}^{A(0)} e^{-jkd} \left(\kappa_{B1}^{B(0)}\right)^*, \tag{70c}$$

$$\kappa_{A1}^{B(0)} = e^{-jk(d_A+d)} \kappa_{B1}^{B(0)}, \tag{70d}$$

$$\mu_{BA}^r = -\kappa_{B1}^{B(0)} e^{-jk(d_A+d)} \left(\kappa_{A1}^{A(0)}\right)^*. \tag{70e}$$

The red arrows combination in Figure 5(c) describes the signal flow dynamics for $A^{bl}$, i.e., resonator A loaded with the resonator B's background structure. The CM equation for $A^{bl}$ is similar to Eq. (69a) but without the term $\mu_{AB}\beta$. Resonator $A^{bl}$ can be considered a mixture of resonator A and the Fabry-Perot-like cavity between resonator A and the ground-plane-like background structure of resonator B. Unlike resonator A, resonator $A^{bl}$ is a one-port resonator as the ground plane blocks the incident waves from the bottom port. Furthermore, from Eq. (69a), the characteristic equation of $A^{bl}$ is $\left|j\omega - P_A^{(bl)}\right| = 0$, which has multiple roots in the complex plane since $P_A^{(bl)}$ is frequency-dependent. Hence, unlike the unloaded resonator A's mode, the mode of resonator $A^{bl}$ has multiple complex resonance frequencies. In contrast to resonator A, resonator B does not experience any background loading effect as resonator A has a free-space background, in agreement with the similarity between the CM equations for resonators B (Eq. (65a)) and $B^{bl}$ (Eq. (69b) without the term $\mu_{BA}\alpha$), as well as the SFGs for resonators B and $B^{bl}$ (blue arrow sets in Figures 5(b) and 5(c), respectively). Only, the top reference plane has shifted from B1 in resonator B to A1 in resonator $B^{bl}$. By decoupling Eqs. (69) via simple mathematical analysis, we can obtain the mode amplitudes as follows:

$$\alpha = \frac{\kappa_{A1}^{A(dl)} S_{inc}}{j\omega - P_A^{(dl)}}, \tag{71a}$$

$$\beta = \frac{\kappa_{A1}^{B(dl)} S_{inc}}{j\omega - P_B^{(dl)}}, \tag{71b}$$

Where we have defined $P_A^{(dl)} = P_A^{(bl)} + \mu_{AB}\mu_{BA}/(j\omega - P_B^{(0)})$, $\kappa_{A1}^{A(dl)} = \kappa_{A1}^{A(bl)} + \mu_{AB}\kappa_{A1}^{B(0)}/(j\omega - P_B^{(0)})$, $P_B^{(dl)} = P_B^{(0)} + \mu_{BA}\mu_{AB}/(j\omega - P_A^{(bl)})$, and $\kappa_{A1}^{B(dl)} = \kappa_{A1}^{B(0)} + \mu_{BA}\kappa_{A1}^{A(bl)}/(j\omega - P_A^{(bl)})$. After that, from Eqs. (68), or alternatively from Eqs. (51) and (61), one can deduce the equations for the output (reflected and transmitted) power waves of the coupled system:

$$\begin{bmatrix} S_R \\ S_T \end{bmatrix} = \mathbf{S^b} \begin{bmatrix} S_{inc} \\ 0 \end{bmatrix} + \mathbf{F} \begin{bmatrix} \alpha \\ \beta \end{bmatrix} = \begin{bmatrix} -e^{-jk(2d_A+2d)} & 0 \\ 0 & -1 \end{bmatrix} \begin{bmatrix} S_{inc} \\ 0 \end{bmatrix} + \begin{bmatrix} e^{-jk(2d_A+2d)}\left(\kappa_{A1}^{A(0)}\right)^* - e^{-jkd_A}\left(\kappa_{A2}^{A(0)}\right)^* & e^{-jk(d_A+d)}\left(\kappa_{B1}^{B(0)}\right)^* \\ 0 & \left(\kappa_{B2}^{B(0)}\right)^* \end{bmatrix} \begin{bmatrix} \alpha \\ \beta \end{bmatrix}. \tag{72}$$

By considering $\mathbf{\Gamma^r} = \begin{bmatrix} \Gamma_A^{r,1}+\Gamma_A^{r,2} & 0 \\ 0 & \Gamma_B^{r,1}+\Gamma_B^{r,2} \end{bmatrix}$ and $\mathbf{\mu^r} = \begin{bmatrix} \mu_{AA}^r & \mu_{AB}^r \\ \mu_{BA}^r & 0 \end{bmatrix}$, Eqs. (64c), (65c), (70), and (72) reveal that $\mathbf{\mu^r} + \left(\mathbf{\mu^r}\right)^\dagger = 2\mathbf{\Gamma^r} - \mathbf{F^\dagger F} \neq 0$ which is compatible with BBCMT formulation of coupled resonators (Eq. (56a)), but beyond TCMT analysis where coupled resonators are assumed closed and radiative coupling is not considered [22, 29].

After calculation of the output power waves From Eq. (72), the system reflected and transmitted powers can be obtained from $|S_R|^2$ and $|S_T|^2$, respectively. Furthermore, from Eqs. (62), (64), and (65), one can get the following equation for the total absorption of the coupled system:

$$A = 2\Gamma_A^{nr}|\alpha|^2 + 2\Gamma_B^{nr}|\beta|^2. \tag{73}$$

From Eq. (73), it is possible to calculate the individual contributions of resonators A and B to the total absorption from $2\Gamma_A^{nr}|\alpha|^2$ and $2\Gamma_B^{nr}|\beta|^2$, respectively.

### 4.2.2. Reverse-engineering the parameters in the BBCMT model

As the next step, we use the kappa-estimation lemma (section 2.6) to reverse-engineer the parameters in the CM equations of the uncoupled resonators by fitting the background-subtracted FDTD-simulated absorbance curves for top-side and bottom-side illuminations. We find the following parameters for the nanodome array (Figures 6(a-b)): $P_A^{(0)} = (1.83\times10^{15}j - 8.10\times10^{13})\,1/s$, $\kappa_{A1}^{A(0)} = 6.05\times10^6\,1/s$ and $\kappa_{A2}^{A(0)} = -8.30\times10^6\,1/s$. Also, we find the nanohole array parameters as follows (Figures 6(c-d)): $P_B^{(0)} = (2.03\times10^{15}j - 3.31\times10^{13})\,1/s$, $\kappa_{B1}^{B(0)} = 1.60\times10^6\,1/s$ and $\kappa_{B2}^{B(0)} = -8.19\times10^5\,1/s$. More details can be found in SI section 4.2 [35]. Interestingly, we find that $\left|\kappa_{A1}^{A(0)}\right| \neq \left|\kappa_{A2}^{A(0)}\right|$ and $\left|\kappa_{B1}^{B(0)}\right| \neq \left|\kappa_{B2}^{B(0)}\right|$, which aligns with the bianisotropic behaviors of the resonators observed in Figures 6(a) and 6(c). From Eq. (18), we can identify the synergistic factors of material loss and non-symmetric resonator structure as the root causes of bianisotropy.

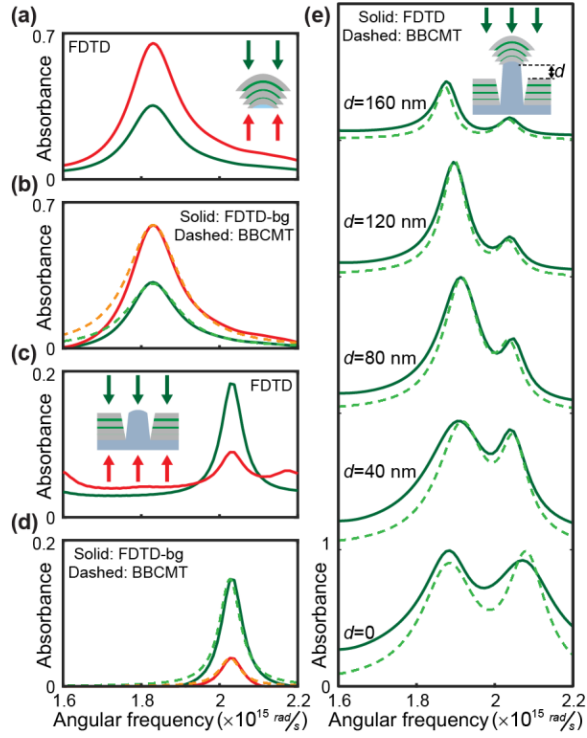

**Figure 6.** Reverse-engineering the parameters in the BBCMT model of the nanolaminate plasmonic crystal from the FDTD-simulated absorbance spectra. (a-d) finding the parameters of the uncoupled resonators by the Kappa-estimation lemma. We used the FDTD-simulated absorbance spectra of the nanodome array (a) and nanohole array (c) for top-side and bottom-side illuminations as the data for the Kappa-estimation lemma. The parameters obtained from the lemma are such that the BBCMT-calculated absorbance spectra for top-side and bottom-side illuminations can fit the background-subtracted absorbance spectra from the FDTD simulations, as illustrated for the nanodome array (b) and the nanohole array (d). (e) We found the parameters of coupling between the resonators by fitting the absorbance spectra of the nanolaminate plasmonic crystal for $d=80$ nm and $d=120$ nm. Nevertheless, the BBCMT-calculated spectra for several other values of $d$ also show a decent match to the FDTD-simulated spectra.

Eqs. (70) gives the background-loaded input-coupling parameters and the coefficients of radiative coupling between the two resonators in terms of the uncoupled resonators' parameters. Nevertheless, we find the near-field coupling parameters in Eqs. (67) by fitting the FDTD-simulated absorbance spectra of the nanolaminate plasmonic crystal for $d=80$ and $d=120$ nm (Figure 6(e)). The fitting process can be simplified by utilizing the Eq. (56b) approximation for the near-field coupling matrix. First, the equation $\mu_{AA}^n + \left(\mu_{AA}^n\right)^* = 0$ indicates that $\mu_{AA}^{n0}$ is a purely imaginary number. Second, the equation $\mu_{AB}^n + \left(\mu_{BA}^n\right)^* = 0$ leads to the relations $\mu_{BA}^{n0} = -\left(\mu_{AB}^{n0}\right)^*$ and $\alpha_{BA} = \alpha_{AB}$. Then, the fitting gives us: $\mu_{AA}^{n0} = 5.7 \times 10^{13} e^{j\pi/2}$ 1/s, $\mu_{AB}^{n0} = -\left(\mu_{BA}^{n0}\right)^* = 8.7 \times 10^{13} e^{j\pi/2}$ 1/s, $\alpha_{AA} = 0.5$ and $\alpha_{AB} = \alpha_{BA} = 2.9$.

Now, by putting the estimated parameters back into Eqs. (71)-(73) and considering $S_{inc} = 1$, we can obtain the (normalized) reflectance, transmittance, and absorbance spectra of the nanolaminate plasmonic crystal for different values of $d$. Interestingly, Figure 6(e) reveals that the BBCMT-calculated absorbance spectra for several values of *d* decently match the FDTD-simulated absorbance spectra, although we only used two values of *d* (*d*=80 and *d*=120 nm) to retrieve the coupling coefficients. This observation demonstrates that the BBCMT model accurately captures the microscopic dynamics in the plasmonic crystal system.

### 4.2.3. Finding the optimal *d* value

For each *d* value, we define the figure of merit for the absorption performance (FOM$_{abs}$) as the area under the absorbance curve in the range $\omega = 1.6 \times 10^{15}$ rad/s to $\omega = 2.2 \times 10^{15}$ rad/s. Figures 7(a-c) represent the absorbance spectra of the nanolaminate plasmonic crystal and each of its subsystems' contributions for *d*=0 up to *d*=2000 nm as colormaps in the $d-\omega$ plane. The corresponding FDTD-simulated colormaps are shown in Figure 7(d-f). Based on the curves of absorption performance extracted from the colormaps in Figure 7 (Figure S3(a-c) [35]), both BBCMT-calculated and FDTD-simulated results indicate that the plasmonic crystal and the nanohole array subsystem show the highest absorption performance for *d*=0. The nanodome array subsystem also shows about 75% of its maximum performance at *d*=0. The BBCMT-calculated and FDTD-simulated colormaps in Figure 7 share a plethora of other similar features as well. For instance: 1) For *d*=0, the peak absorbance of the nanodome array is 2.5 times that of the uncoupled nanodome array for excitation from the top port, while this ratio is 4.5 for the nanohole array; 2) As *d* goes up, both the peak absorbance and FOM$_{abs}$ of the nanohole array considerably drop and do not recover again, but this is not the case for the nanodome array; 3) The absorbance spectra for the plasmonic crystal and both of its sub-arrays asymptotically approach a periodic function of *d* with the period being half of the wavelength ($\lambda/2 = \pi c/\omega$) at each frequency (Figure S3(d-f) [35]).

Unlike FDTD, BBCMT offers a microscopic picture that helps us understand the enabling factors behind the described features. The term $-\kappa_{A2}^{A(0)} e^{-jk(d_A + 2d)}$ in Eq. (70b) introduces an indirect channel for coupling of the nanodome array mode to the incident light through interaction with the non-resonantly backscattered light from the nanohole array. In addition, magnetic dipole and bianisotropic natures of the nanodome array mode cause $\text{sign}(\kappa_{A1}^{A(0)}) \neq \text{sign}(\kappa_{A2}^{A(0)})$ and $|\kappa_{A2}^{A(0)}| > |\kappa_{A1}^{A(0)}|$, as shown in section 4.2.2, respectively. These factors make $\kappa_{A1}^{A(dl)}$ in Eq. (71a) much larger than the intrinsic $\kappa_{A1}^{A(0)}$ value in a broad frequency range when *d*~0 (Figure S4(b) [35]), leading to an enhanced absorbance for the nanodome array in this regime (Figure S4(a) [35]). Yet, the bandwidth of this enhancement is limited by the resonance factor $1/(j\omega - P_A^{(dl)})$ (Figure S4(c) [35]). From Eq. (71b), high absorbance enhancement of the nanohole array for *d*~0 (Figure S4(d) [35]) is due to the strong coupling between the two resonators (large $\mu_{AB}$ and $\mu_{BA}$) in this regime, which enhances $\kappa_{A1}^{B(dl)}$ much beyond the intrinsic $\kappa_{A1}^{B(0)}$ value (Figure S4(e) [35]). The near-field coupling terms $\mu_{AB}^n$ and $\mu_{BA}^n$ have the dominant contribution to the resonators' strong coupling for *d*~0. As *d* goes up, $\mu_{AB}^n$ and $\mu_{BA}^n$ exponentially decay, leading to a conspicuous absorption reduction in the nanohole array. Moreover, from Eqs. (70) and (71), the mode amplitudes $\alpha$ and $\beta$ become periodic functions of *d* with the periodicity of $\lambda/2$ at each frequency. From Eq. (73), the mode amplitudes' periodic behavior for large *d* values is the reason for the periodicity of the absorbance spectra for the plasmonic crystal and its subarrays in this regime. The microscopic picture offered by the BBCMT model allows us to definitively conclude that the nanolaminate plasmonic crystal has the highest absorption performance for *d*=0. In addition to absorbance, we have calculated the reflectance and transmittance spectra of the plasmonic crystal for *d*=0 up to *d*=2000 nm from Eq. (72) (Figure S5 [35]). The resulting $d-\omega$ plane colormaps again show a high similarity level to the FDTD-calculated reflectance and transmittance colormaps (SI section 4.2 [35]).

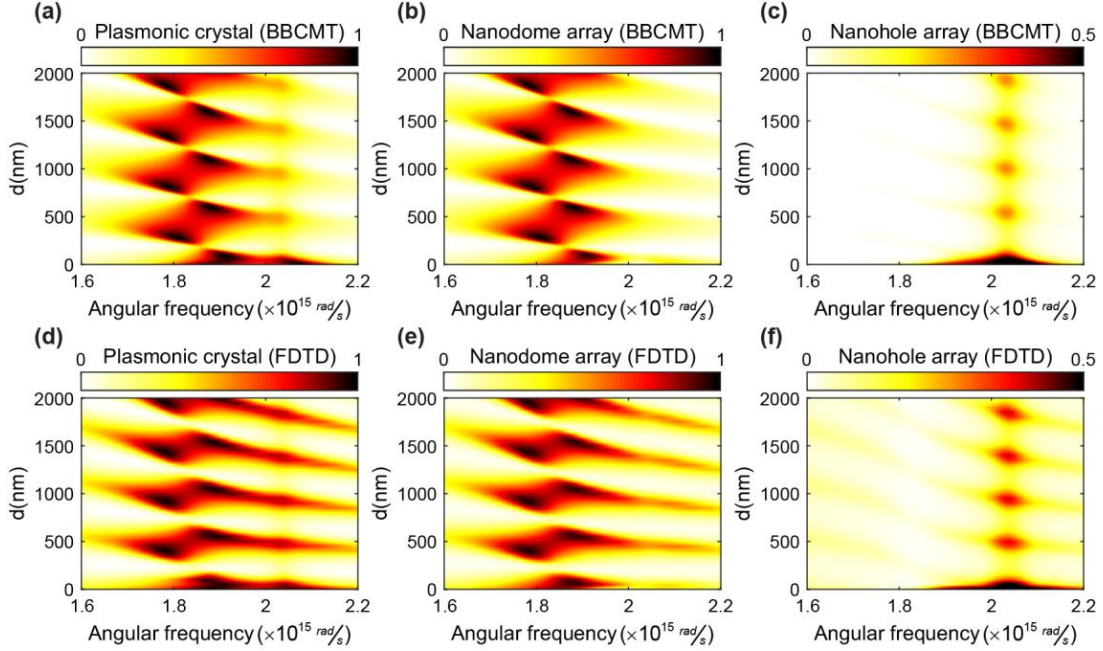

**Figure 7.** Side-by-side comparison between the BBCMT calculations and FDTD simulations for the absorbance spectra of the nanolaminate plasmonic crystal and its sub-arrays for $d=0$ up to $d=2000$ nm. The resulting spectra are shown as colormaps in the $d-\omega$ plane. (a, d) Absorbance colormap of the nanolaminate plasmonic crystal from BBCMT (a) and FDTD (d). (b, e) Absorbance colormap of the coupled nanodome array from BBCMT (b) and FDTD (e). (c, f) Absorbance colormap of the coupled nanohole array from BBCMT (c) and FDTD (f).

There still exist small discrepancies between the BBCMT and FDTD results, such as the spectra's linewidth for different $d$ values. A prime source of these discrepancies is the relative inaccuracy of assuming a free-space background for the nanodome array as it neglects the broadband scattering from the electric dipole mode of the nanodome array [38, 40]. A more accurate option for the background structure of the nanodome array is a similar nanodome array but with the gaps filled with metals instead of dielectrics (like section 4.1). Such background structure supports almost a similar electric dipole mode as the original nanodome array but does not support the magnetic modes associated with the gaps.

As the final remark on this part, we examined the importance of considering the non-resonant loading effect via eliminating it from the model, i.e., putting $P_A^{(bl)} = P_A^{(0)}$ in Eqs. (69), and then comparing the resulting far-field spectra for different $d$ values (Figure S6 [35]) against the FDTD- calculated spectra. We have re-adjusted the mode-coupling coefficients ( $\mu_{AB}^{n0} = -\left(\mu_{BA}^{n0}\right)^* = 2\times 10^{13} e^{(j7\pi/8)}$ 1/s ) to fit the absorption spectrum of the nanolaminate plasmonic crystal for $d=80$ nm as best as possible. Nevertheless, as indicated in Figure S6 [35], the results predicted from this simplified model are still highly off from those calculated by FDTD, revealing that accounting for the non-resonant loading effect is vital for accurate modeling of the system. The non-resonant loading effect is not considered in the TCMT framework.

## 5. Summary and Outlook

In summary, we presented the BBCMT, a general ab initio CMT framework developed from Maxwell's equations and QNM theory that can model electromagnetic interactions of open, lossy, and dispersive resonators. The BBCMT framework's development is enabled by our novel QNM analysis approach that by using the conjugated Lorentz's reciprocity and Poynting's theorems, rigorously normalizes QNMs to define a resonator Hamiltonian matrix. The Hamiltonian matrix characterizes the resonator modes' electromagnetic interactions, energy storage, and absorption. To approximate a resonator's non-resonant scattering and absorption responses, the framework can treat the resonator scattered fields as a perturbation to those of a user-defined background structure. The background responses also capture the responses of the modes not associated with the user's principal areas of interest in the resonator. Choice of the background structure is flexible and depends on the desired levels of accuracy and complexity. BBCMT modeling complexity can further be adjusted by treating user-defined subcomponents of a resonator system as black-boxes described only by their input-out transfer characteristics. Besides the BBCMT formulation, we offered the Kappa-estimation and F-estimation lemmas to retrieve the modeling parameters from the calculated or measured far-field spectra. Furthermore, we introduced the signal flow graphs from control engineering to illustrate and interpret the electromagnetic interactions of resonator systems and solve their corresponding BBCMT equations using Mason's theorem.

To demonstrate the validity, generality, and flexibility of BBCMT, we discussed two examples where the predictions of BBCMT and TCMT models were closely compared to FDTD simulations and found the BBCMT results to be a much better match. Furthermore, we showed that BBCMT can reduce to TCMT and cavity perturbation theory under certain conditions and approximations. Our work demonstrates that BBCMT is a versatile theoretical platform to facilitate the design and analysis of linear resonator systems across the whole electromagnetic spectrum, especially for plasmonics and metamaterials applications. As future work, BBCMT can be further generalized to handle parametric and non-parametric nonlinearities as well.